\documentclass[twocolumn,showpacs,prl,preprintnumbers,amsmath,amssymb]{revtex4}

\usepackage{graphicx}
\usepackage{dcolumn}
\usepackage{bm}
\usepackage[normalem]{ulem}
\usepackage{color}
\usepackage{amssymb}
\usepackage{amstext}

\begin{document}

\title{Optical response of ferromagnetic YTiO$_3$ studied by spectral ellipsometry}

\author{N.N. Kovaleva}
\author{A.V. Boris}
\author{P. Yordanov}
\author{A. Maljuk}
\author{E. Br\"{u}chner}
\author{J. Strempfer}
\author{M. Konuma}
\author{I. Zegkinoglou}
\affiliation{Max-Planck-Institut f\"{u}r Festk\"{o}rperforschung, Heisenbergstrasse 1, D-70569 Stuttgart, Germany}
\author{C. Bernhard}
\affiliation{Department of Physics, University of Fribourg, Chemin du Mus\'{e}e 3, CH-1700, Fribourg, Switzerland}
\author{A. M. Stoneham}
\affiliation{Department of Physics and Astronomy, University College London,
Gower Street, London WC1E 6BT, United Kingdom}
\author{B. Keimer}
\affiliation{Max-Planck-Institut f\"{u}r Festk\"{o}rperforschung, Heisenbergstrasse 1, D-70569 Stuttgart, Germany}

\date{\today}

\begin{abstract}
 
We have studied the temperature dependence of spectroscopic ellipsometry
spectra of an electrically insulating, nearly stoichiometric
YTiO$_{3}$ single crystal with ferromagnetic Curie temperature $T_C$
= 30 K. The optical response exhibits a weak but noticeable
anisotropy. Using a classical dispersion analysis, we identify three
low-energy optical bands at 2.0, 2.9, and 3.7 eV. Although the
optical conductivity spectra are only weakly temperature dependent
below 300 K, we are able to distinguish high- and low-temperature
regimes with a distinct crossover point around 100 K. The
low-temperature regime in the optical response coincides with the
temperature range in which significant deviations from Curie-Weiss
mean field behavior are observed in the magnetization.  Using an analysis based on a simple superexchange model, the spectral weight rearrangement can be attributed to intersite $d_i^1d_j^1$ $\longrightarrow$ $d_i^2d_j^0$ optical transitions. In particular, Kramers-Kronig consistent changes in
optical spectra around 2.9 eV can be associated with the high-spin-state ($^3T_1$) optical transition.
This indicates that other mechanisms, such as weakly dipole-allowed $p-d$ transitions and/or exciton-polaron excitations, can contribute significantly to the optical band at 2 eV. The recorded optical spectral weight  gain of 2.9 eV optical band is significantly suppressed and anisotropic, which we associate with complex $spin-orbit-lattice$ phenomena near ferromagnetic ordering temperature in YTiO$_3$.     

\end{abstract}


\date{\today}

\maketitle

\section*{I. INTRODUCTION} 

The relationship between the orbital state and the physical
properties of YTiO$_3$, along with its antiferromagnetic partner
LaTiO$_3$, has generated significant recent attention. Electronic
structure calculations predict a specific, four-sublattice
arrangement of $t_{2g}$ orbitals supporting ferromagnetic order in
YTiO$_3$ \cite{Mochizuki,Pavarini,Schmitz,Solovyev,Mizokawa}. Direct
evidence for an orbitally ordered state akin to that predicted in
theoretical work has come from neutron diffraction \cite{Akimitsu}.
X-ray absorption \cite{Tjeng} and nuclear quadrupole resonance
studies \cite{Itoh} have yielded supporting evidence. However, some
major questions remain open. While electronic-structure calculations
generically predict a strong spatial anisotropy of the superexchange
parameters, inelastic neutron scattering experiments have revealed
an almost isotropic spin wave spectrum \cite{Ulrich}. The spectrum
of orbital excitations recently revealed by Raman scattering
is also in stark contrast to predictions based on the
theoretically proposed orbitally ordered state \cite{Ulrich06}.
Collective quantum zero-point fluctuations that strongly reduce the
orbital order might offer a way out of this conundrum, but
theoretical work on this scenario is still in its early stages
\cite{Khaliullin}.

Here we present a comprehensive
ellipsometric study of the optical properties of stoichiometric
YTiO$_3$, following the same methodology as in previous work on
LaMnO$_3$ \cite{Kovaleva}. In LaMnO$_3$, the Mott-insulating progenitor of a family of compounds exhibiting colossal magnetoresistance, a realistic quantitative description of the interplay between
spin and charge correlations has found possible \cite{Kovaleva}. The four
$d$-electrons of LaMnO$_3$ are subject to a nearly cubic crystal
field that splits the energy levels available to these electrons
into a higher-lying, doubly degenerate $e_g$ state and a
lower-lying, triply degenerate $t_{2g}$ state. Following Hund's
rules, these states are occupied by one and three electrons,
respectively, so that the $e_g$ valence electron occupies a doubly
degenerate level. This degeneracy is lifted by a cooperative
Jahn-Teller distortion (``orbital order") around 800 K. The
anisotropic charge distribution that goes along with orbital order
gives rise to optical birefringence. In particular, the spectral
weight of the Mott gap excitations around 2 eV becomes anisotropic
below 800 K \cite{Tobe}. Ellipsometric measurements revealed that
this anisotropy increases strongly below the antiferromagnetic
ordering temperature of $\sim 150$ K \cite{Kovaleva}. The dependence
of the optical spectra on the alignment of nearest-neighbor Mn spins
reveals that the lowest-energy excitations are intersite $d-d$
transitions. A quantitative description of the temperature
dependence of the optical spectral weight could be achieved, based on the realization that the same fundamental mechanisms that determine the spectral weight of the optical excitations (namely, the Pauli principle and Hund's rules) also underlie the superexchange interaction parameters between Mn
spins that had been obtained independently by inelastic neutron
scattering.  

As YTiO$_3$ is isostructural to LaMnO$_3$, but possesses only a
single valence $d$-electron, it is not unreasonable to expect that
its optical properties are amenable to a description in the same
framework. However, there are several important differences between
both compounds. YTiO$_3$ is {\it ferromagnetic} below $T_C \sim 30$ K
\cite{Ulrich} and it is hence an uncommon example of an electrically
insulating ferromagnet. Moreover, YTiO$_3$ does not exhibit any
structural phase transition below its chemical decomposition
temperature of around 700 K \cite{Cwik}. A cooperative Jahn-Teller
effect akin to that in LaMnO$_3$ is thus not apparent in its phase
behavior. These differences are rooted in the microscopic properties
of the orbitals occupied by the valence electrons. Specifically, the
$d$-electron of YTiO$_3$ occupies the $t_{2g}$ manifold, which
exhibits a higher degeneracy and weaker lattice coupling than the
$e_g$ electron in LaMnO$_3$.

The comprehensive optical spectroscopy study of carefully prepared
single crystals of YTiO$_3$ reported here was designed to motivate
and guide further theoretical work on the interplay between spin and
charge correlations in the titanates. The experiments proved
surprisingly difficult, because the optical spectra were found to
depend strongly on the type of surface exposed to the beam. Through
extensive characterization measurements, this effect could be traced
to an accumulation of oxygen defects near some (but not all)
surfaces of the perovskite structure. Special care has to be taken
in order to maintain stoichiometry in the crystal volume probed by
the beam. After these difficulties were overcome, the results
revealed a situation that is strikingly different from that in
LaMnO$_3$. Notably, the optical spectrum depends only weakly on
light polarization and on temperature, and the spectral weight
rearrangement at the Curie temperature is extremely small. A
noticeable spectral weight rearrangement is, however, observed at
about 100 K, far above the Curie temperature. Possible origins of
this effect are discussed.

\section*{II. CRYSTAL GROWTH AND CHARACTERIZATION}

\subsection*{A. Crystal growth}
Single crystals of YTiO$_3$ were grown by the floating
zone method in a reducing atmosphere (Ar/H$_2$ = 50/50). The
starting material was prepared by mixing appropriate amounts of
Y$_2$O$_3$ (5N) and Ti$_2$O$_3$ (3N) powders in ethanol (99.8\%
purity). Note that commercially available Y$_2$O$_3$ powders
contain 0.6 - 0.8 wt.\% of water. Ignoring this during the feed
rod preparation leads either to a deficiency of yttrium, or to
titanium oxide impurities in the as-grown crystal. The mixture was
calcined at 950$^{\circ}C$ in Al$_2$O$_3$ crucibles for 12h in
vacuum (10$^{-6}$ mbar). The calcined powder was formed into a
cylindrical shape of 7-8 mm in diameter and 100 mm in length, and
pressed at a hydrostatic pressure of about 400 MPa. The rods were
sintered at 1050$^\circ$C for 12 h under the same vacuum
conditions. The apparatus used for crystal growth was a
four-mirror-type infrared image furnace (Crystal System Corp.,
FZ-T-10000-H-III-VPR) equipped with four 1.5 kW halogen lamps. In
the growth process the seed and feed shafts were rotated in
opposite directions at rates of 10-15 rpm, the pulling rate was
varied from 5 to 10 mm/h. The grown ingots were cut into wafers
perpendicular to the growth direction, and both sides of wafers
were polished to mirror finish. Then polished specimens were
characterized using a polarizing optical microscope to examine the
presence of sub-grains and inclusions.

\subsection*{B. Structure and oxygen stoichiometry}
YTiO$_3$ exhibits the orthorhombic structure of GdFeO$_3$ type of
$Pbnm$ space group (see Fig. 1). X-ray diffraction measurements
show that the YTiO$_3$ single crystals used for our experiments
are untwinned, with mosaicity less than 0.03$^\circ$. At room
temperature, the lattice parameters are $a$ = 5.331(3), $b$ =
5.672(4), and $c$ = 7.602(6) $\AA$, slightly different from those
of Czochralski-grown single crystals reported by Maclean {\it et
al.} \cite{Maclean}. The samples were aligned along the principal
axes and cut in the form of a parallelepiped with dimensions $\sim
3 \times 3 \times 3$ mm$^3$.

The oxygen content was measured by heating small quantities of the
as-grown crystals in O$_2$ flow using a DTA-TG apparatus.
Due to the Ti$^{3+}$ $\rightarrow$ Ti$^{4+}$ instability, single
crystals of YTiO$_3$ always have an excess of oxygen above the
exact stoichiometry formula. Based on the weight gain, the oxygen
excess in the formula YTiO$_{3+\delta}$ is estimated at a level
less than $\delta$ = 0.013 (see Fig. 2). As follows from the
figure, Ti$^{3+}$ ions fully oxidize to Ti$^{4+}$ above 300
$^\circ$C in flowing oxygen. It is important to mention that
oxidation takes place above 450 $^\circ$C in air. Therefore, one
should avoid any warming above this temperature during the cutting
and polishing procedures.

\subsection*{C. Magnetization}

The samples were further characterized by magnetometry, using a
superconducting quantum interference device. The inset of Fig. 3
shows temperature dependence of the magnetization for YTiO$_3$
sample in the vicinity of T$_C$ taken on heating in a magnetic
field of 1 kOe parallel to the $c$-axis direction, after cooling
in zero magnetic field. For the almost stoichiometric sample
studied here, we estimate T$_C$ =
30 K, amongst the highest values reported so far for YTiO$_3$.

In Fig. 3 we show the temperature dependence of the inverse
susceptibility, $1/\chi_{mol}$, which exhibits a nearly linear increase above 100 K ($\chi_{mol}$ was corrected by a temperature-independent diamagnetic contribution
$\chi_{dia}$ = - 5.7 $\times$ 10$^{-5}$ emu/mol \cite{Itoh}). A
fit to a Curie-Weiss law, $1/\chi_{mol} \approx \frac
8{\mu_{eff}^2}(T-\theta)$, yields a Curie-Weiss temperature
$\theta$ = 40 K and an effective magnetic moment $\mu_{eff}$ =
1.66 $\mu_{B}$, which is slightly reduced with respect to the
spin-only value $\mu_S = 2\sqrt{S(S+1)}$
 = 1.73 $\mu_{B}$ of the Ti$^{3+}$ ions with S = 1/2. The magnetization data of Fig. 3 reveal significant deviations from Curie-Weiss mean-field behavior below about 100 K. As a result, an inflection point in the temperature-dependent magnetization is observed at 27 K, several degrees below $T_C$. 

Figure 4 shows magnetization measurements for the three principal
directions of YTiO$_3$ up to 7 T at 5 K, well below the magnetic
transition temperature. These magnetic measurements reveal that the $c$-axis is the easy axis of magnetization, while
the $b$-axis is hard. The saturated moment, which we estimate in
the easy direction  at the highest magnetic field of 7 T is
$\mu^{c}$ = 0.84 $\mu_B$, whereas $\mu^{b}$ = 0.82 $\mu_B$ in the
hard direction, both are   reduced from one Bohr magneton. The data are in
good agreement with those reported by Tsubota {\it et al.}
\cite{Tsubota} on single crystals also grown by a floating-zone
method. By contrast, earlier data on single crystals 
grown by the Czochralski technique showed markedly reduced
magnetic fields at which the saturation behavior of the
magnetization sets on \cite{Garrett}. The anisotropy of the
magnetization is also different: in the Czochralski-grown
crystals, both the $a$- and $c$-axes were established to be easy
axes.

\section*{III. RESULTS AND DISCUSSION}

\subsection*{A. Technical details}

For optical measurements the surfaces were polished to optical
grade. The measurements in the frequency range of 4000-44000
cm$^{-1}$ (0.5-5.5 eV) were performed with a home-built ellipsometer of
rotating-analyzer type \cite{Kircher}, where the angle of
incidence is 70.0$^\circ$. The sample was mounted on the cold
finger of a helium flow UHV cryostat with a base pressure of
5x10$^{-9}$ Torr at room temperature.  In the spectral range
80-6000 cm$^{-1}$ (0.01-0.75 eV) we used home-built ellipsometers in combination
with a fast-Fourier transform interferometer
at the infrared beam line of the ANKA synchrotron at the
Forschungszentrum Karlsruhe, Germany \cite{Bernhard}. In addition,
near-normal incident reflectivity was measured in the phonon frequency
range of 80-700 cm$^{-1}$ using a Fourier transform spectrometer Bruker IFS66vs.

\subsection*{B. Infrared-active phonons}

We now present a full set of spectra of infrared-active optical
phonons. YTiO$_3$ has an orthorhombic crystal structure (space
group {\it Pbnm}, D$^{16}_{2h}$) with four formula units per unit
cell (see Fig. 1). A factor-group analysis yields a total number
of 60 $\Gamma$-point phonons, of which 24
[7A$_g$+5B$_{1g}$+7B$_{2g}$+5B$_{3g}$] are Raman-active modes, 25
[9B$_{1u}$+7B$_{2u}$+9B$_{3u}$] are infrared-active modes, 8 A$_u$
are silent modes, and 3[B$_{1u}$+B$_{2u}$+B$_{3u}$] are acoustic
modes. We performed spectroscopic ellipsometry on the $ac$- and $bc$-surfaces
for all high-symmetry orientations
of the optical axis with respect to the plane of incidence of light and near-normal
incidence reflectivity measurements for {\bf E} $\mid\mid$ {\bf a}, {\bf E} $\mid\mid$ {\bf b}, and {\bf E} $\mid\mid$ {\bf c}.

In Fig. 5 we show low-temperature optical conductivity spectra in
the phonon frequency range. The spectra were deduced from
polarized reflectivity spectra via a standard Kramers-Kronig
analysis, combined with the respective ellipsometry spectra for
correct normalization. Our polarized optical measurements allowed
us to observe fully anisotropic phonon spectra in YTiO$_3$ single
crystals and make an assignment of the infrared-active modes
according to their symmetry. For {\bf E} $\mid\mid$ {\bf a} we
distinguish 9B$_{1u}$ normal modes at the frequencies 154, 210,
316, 336, 428, 440, 516, 554, and 577 cm$^{-1}$, for {\bf E}
$\mid\mid$ {\bf b} we distinguish 9B$_{3u}$ normal modes at the
frequencies 139, 240, 273, 309, 364, 380, 463, 523, and 578
cm$^{-1}$, and for {\bf E} $\mid\mid$ {\bf c} we distinguish
7B$_{2u}$ normal modes at the frequencies 165, 203, 322, 349, 386,
543, and 563 cm$^{-1}$. 
Our observation of fully polarized phonon spectra and its consistency with the symmetry analysis indicate
good alignment of the crystal with respect to the crystallographic
directions. Further, the narrow and intense phonon peaks are
testimony to the excellent quality of our YTiO$_3$ crystal.

\subsection*{C. Surface defect states}

We have found out that polarized optical spectra in YTiO$_3$ exhibit specific features associated  with the surface type exposed to the beam. Figure 6 shows spectra of high-frequency, oxygen-dominated optical phonons measured with light polarized along the $b$-axis on different types of the surface associated with the $ab$ and $bc$ crystallographic planes. Although the main features are observed at the same frequencies in both cases, a number of side-band modes are present in measurements on the $ab$ surface. A similar set of extra modes is observed in $a$-axis polarized measurements on the $ab$ surface, compared to the $ac$ surface. As single crystals of YTiO$_{3+\delta}$ easily accommodate extra oxygen, the appearance of the weaker satellites in the phonon spectra can be naturally explained as arising from the coupling with   local oxygen modes, associated with  oxygen interstitials, predominating in the $ab$ surface. 

In addition, we show that the optical conductivity spectra close
to the absorption edge are also dependent on the type of surface. In the inset of Fig. 7 one can see that the optical conductivity spectrum measured along the $b$-axis on the $ab$ surface exhibits an optical band at about 0.8 eV, near the fundamental absorption edge. As a result, the absorption edge shifts to lower energies, compared to the $bc$ surface. This may
be an indication on the filling of the gap states with the energy
levels associated with the defects predominating in the $ab$
crystallographic plane. It is well known that the absorption
within the optical gap in the RTiO$_3$ system is extremely
sensitive to the oxygen stoichiometry \cite{Okimoto,Taguchi}.
Therefore, these data provide additional evidence for the presence
of defects associated with oxygen interstitials in the $ab$
surface, in agreement with the phonon spectra.

Motivated by the optical data, we have performed
surface-characterization measurements using Auger-electron
spectroscopy (AES). Figure 8 shows typical AES sputter-depth
profiles obtained by Ar$^+$ bombardment of different YTiO$_3$
surfaces. One can see that the titanium concentration is almost
flat in the sputter-depth profile, and independent of the type of
surface studied. This is also the case for the oxygen profile
measured on the $bc$ surface of YTiO$_3$ crystals. On the contrary, monitoring the oxygen concentration in the $ab$ surface shows the profile  significantly extended into the sample depth. 

We now discuss the origin of the differences in oxygen defect
concentration near the different types of surfaces. Ignoring the
orthorhombic distortion for simplicity, we note that the $bc$- and
$ab$-surfaces correspond (110) and (001) planes of the simple
cubic perovskite structure, respectively (see Fig. 1). The atomic stacking sequences of these planes are O$_2$-YTiO-O$_2$-YTiO-... and
YO-TiO$_2$-YO-TiO$_2$-..., so that  the valence states of the
constituent ions imply that these surfaces are
polar. Polar surfaces are intrinsically unstable due to the
macroscopic polarization caused by a dipole moment perpendicular
to the surface generated by the alternating, oppositely charged
planes \cite{Noguera}. For the perovskite structure of LaMnO$_3$, it has been shown by $ab$-$initio$ calculations \cite{Evarestov} that the
polarity can be lifted in different ways for different types of surfaces.
For instance, a redistribution of charge is able to cancel the
macroscopic polarization near the (001) surface, whereas surface oxygen vacancies
serve as a better stabilizing factor at the (110) surface. Of course, these calculations cannot be simply transposed to YTiO$_3$, nonetheless, they indicate a possible microscopic origin of the differences in oxygen profiles and optical spectra revealed by our data.

Thus in the case of the $ab$ surface (associated with the (001) plane of the perovskite structure) the charge redistribution could result in a formation of the defects comprising oxygen interstitials in the extended near-surface region. This could effect the infrared oxygen phonon modes and optical absorption edge in YTiO$_3$ crystal, in agreement with our observations. Contrary to the $ab$ surface, the perpendicular surfaces (associated with the
(110) plane of the perovskite structure) could have predominantly surface defects. Accordingly, we clearly see that the optical conductivity spectra
measured on the surfaces including the $c$-axis show a dip at around 4.5 eV, as illustrated by Fig. 7. This effect is by far more pronounced in the optical conductivity spectra and appears at the UV frequencies, close to the onset of the strongly dipole-allowed $p-d$ transition. These observations show the importance of the surface effects,
associated with screening of the surface charges by crystal defects and bond-bending
phenomena at the surface, on the optical conductivity of YTiO$_3$. Therefore, one should refer the polarized optical conductivity spectra to the type of the surface measured. Most of the measurements
discussed below were taken on the $bc$-surface, where no excess
oxygen is detected. 

Aging effects on the optical spectra of samples kept in air for extended periods of time are presumably also related to oxygen incorporation. Nonequilibrium conditions at the crystal surface due  to the Ti$^{3+}$ $\rightarrow$ Ti$^{4+}$ instability in air are  conducive to the formation of the titanium oxides, with the valency of Ti  increased above 3+, and simultaneous incorporation
of extra oxygen (interstitials). Due to the diffusion processes these defects migrate from the surface into the crystal bulk and form an extended  profile.
In Fig. 7, optical conductivity spectra in $b$-axis polarization on freshly polished $ab$-surfaces are compared to identical data taken
after about 1-1/2 years. The flat background, which sets on around 1.9 eV and extends to higher energies, decreases progressively with time.   We show later on that the temperature evolution of the
optical response measured on aged surfaces differs drastically
from the temperature effects we associate with the  bulk
properties of stoichiometric YTiO$_3$.

\subsection*{D. Dielectric response and optical conductivity}

\subsubsection*{{\it 1. Overall description and dispersion analysis}}

Figures 9 and 10 show the anisotropic real and imaginary parts of the
dielectric function, $\tilde \epsilon(\nu)$ = $\epsilon_1(\nu) +
i\epsilon_2(\nu)$, and the optical conductivity
$\sigma_1(\nu)=1/(4\pi)\nu \epsilon_2(\nu)$ at T = 300 K extracted
from ellipsometric data. In contrast to prior
data deduced from reflectivity measurements via Kramers-Kronig analysis \cite{Taguchi,Arima,Okimoto,Katsufuji}
we are able to resolve a noticeable anisotropy in the intensity of the $ab$-plane and $c$-axis dielectric response. We can also distinguish a weak anisotropy within the $ab$-plane. In agreement with prior work \cite{Taguchi,Arima,Okimoto,Katsufuji}, our optical conductivity spectra exhibit a smooth onset of optical absorption around 0.5 eV (Fig. 7). Due to the Mott nature of the optical gap in YTiO$_3$, formed between the lower and upper $d$ bands, the fundamental absorption edge is supposed to exhibit $d-d$ character. However, the origin of the low-energy transitions in optical spectra of YTiO$_3$ has not been addressed in details so far. In our $\epsilon_2(\nu)$ and $\sigma_1(\nu)$ spectra one can clearly see a  lowest-energy optical band peaking around 1.9 eV, whereas the following optical transitions are strongly superimposed, and the optical spectra look like featureless background for the highly intense
band around 5.5 eV. To separate contributions from the different low-energy optical bands, we perform a classical dispersion analysis. Figure 11 summarizes results of a dispersion analysis of the complex dielectric response for $b$-axis polarization measured on $ab$ plane at T = 300 K. Using a dielectric function of the form $\tilde\epsilon(\nu) = \epsilon_{\infty}
+ \sum_j \frac{S_j}{\nu_j^2-\nu^2-i\nu\gamma_j}$, where $\nu_j$, $\gamma_j$,
and $S_j$ are the peak energy, width, and dimensionless oscillator strength
of the $j$th oscillator, and $\epsilon_{\infty}$ is the core contribution
from the dielectric function, we fit a set of Lorenzian oscillators simultaneously to $\epsilon_1(\nu)$ and $\epsilon_2(\nu)$. In our analysis we introduce a minimum set of oscillators, with one oscillator beyond the spectral range investigated. The determined parameters of $\nu_j$, $S_j$, and $\gamma_j$ are listed in Table 1.  In accordance with our dispersion analysis, the three low-energy optical bands are located at  2.0, 2.9, and 3.7 eV in $b$-axis polarization at T = 300 K. An important value is the associated spectral weight (SW), which can be estimated    for the separate Lorentz oscillator as $SW = \int\sigma_1(\nu^{'})d\nu^{'}$ = $\frac{\pi}{120}S_j\nu_j^2$. In Table 1 we list normalized SW values in terms of the effective numbers of electronic charge carriers $N_{eff}$ =$\frac{2m}{\pi e^2N} SW$, where $m$ is the free electron mass and $N$ = $a_0^{-3}$ = $1.7 \times 10^{22}$ cm$^{-3}$ is the density of Ti atoms. Note that the dip in the
optical $c$-axis response above 4 eV is strongly influenced by surface effects,
as discussed in Section III.C above, therefore, the fit is less accurate
in $c$-axis polarization. However, we have estimated
the corresponding $\nu_j$ and $S_j$ values in $c$-axis polarization, while
keeping the widths of the optical bands,$\gamma_j$, the same as in $b$-axis
polarization. These values are listed in Table I in round brackets.      
   
\begin{table}
\caption{Parameters of Lorenz oscillators resulting from dispersion analysis of complex dielectric response in $b$-axis ($c$-axis) polarization in YTiO$_3$ measured on $ab$ plane ($bc$ plane) at T = 300 K, $\epsilon_{\infty} = 1.78$
($\epsilon_{\infty}$ = 2.18).}
\begin{ruledtabular}
\begin{tabular}{llll}
$\nu_j$ (eV) & $S_j$ & $\gamma_j$ (eV)& $N_{eff}$\\   
\colrule
1.95 (2.11) & 1.25 (0.67) &  1.68 & 0.21 (0.13)\\
2.94 (2.78) & 0.20 (0.08) &  1.46 & 0.07 (0.03)\\
3.72 (3.43) & 0.08 (0.12) &  0.96 & 0.05 (0.06)\\
5.05 (5.25) & 0.22 (0.24) &  0.80 & 0.24 (0.29)\\
5.61 (5.78) & 0.84 (0.82) &  1.15 & 1.14 (1.20)
\end{tabular}
\end{ruledtabular} 
\end{table}

Three main contributions have to
be distinguished in the spectral range of interest here: (i) $p-d$ transitions
from the occupied O $2p$-band into the partially occupied Ti $d$-levels; (ii) intra-site crystal-field (CF) $d-d$ transitions from the partially occupied Ti $t_{2g}$ manifold into the empty $e_g$ levels; and (iii) inter-site $d-d$ transitions within the $t_{2g}$ manifold and between  the $t_{2g}$ and $e_g$
levels. In addition, the associated excitonic and polaronic optical bands
may also contribute in the spectral range of interest. Based on comparison to optical data on other transition metal oxides and to a variety of theoretical calculations \cite{Solovyev1,Bouarab,Krasovska}, the pronounced optical band at $\sim$ 5.5 eV can be assigned to strongly dipole-allowed $p-d$ transition. An assignment of the lower-energy bands is, however, much more difficult. In particular, cluster calculations have identified a series of weakly
dipole-allowed charge-transfer (CT) transitions O$2p$ $\longrightarrow$ Ti $3d$ with energies and spectral weights comparable to those observed in our experiment \cite{Zenkov}. These transitions overlap the $d-d$ transitions also expected to be present in this energy range.

\subsubsection*{{\it 2. Temperature dependence}}

Recent work on LaMnO$_3$ has shown that the temperature dependence of the
optical spectral weight can be instrumental in tracking down the origin of the various features in the optical response \cite{Kovaleva}. 
The spectral weight of the inter-site $d-d$ transitions is very sensitive to the temperature dependent spin correlations, because the spin alignment controls the transfer of electrons between neighboring sites via the Pauli principle. The contribution from these transitions can thus be singled out by monitoring the evolution of the optical response through the onset of magnetic order, as demonstrated in LaMnO$_3$ \cite{Kovaleva}. 

Figure 12 shows that the temperature effect on the optical spectra is weak in the temperature range from room temperature down to low temperatures. The main effects
are a shift of the absorption edge to higher energies and a narrowing of the strong optical band at about 5.5 eV. Our ellipsometric measurements also allowed us to resolve much weaker effects at low temperature, which one may associate with redistribution of optical spectral weight around the spin ordering temperature T$_C$ = 30 K. Figures 13a and b show difference in the
dielectric function spectra, $\Delta \epsilon_1(\nu)$ and $\Delta
\epsilon_2(\nu)$, between 15 and 55\ K, measured along the $b$- and
$c$-axis, respectively. The $b$-axis difference spectra exhibit a weak
resonance feature in $\Delta \epsilon_2(\nu)$ around 2.85 eV and
an antiresonance feature (zero-crossing) in $\Delta
\epsilon_1(\nu)$ at the same energy. The observed changes in
$\tilde \epsilon(\nu)$ provide experimental evidence for the
presence of an optical band centered around 2.85 eV. The intensity of this band increases below the Curie temperature. Similar reasoning
shows that the intensity of the optical band around 3.95 eV 
decreases below T$_C$. By contrast, the optical response in $c$-axis
polarization is temperature independent in this spectral range, to within our experimental accuracy (Fig. 13b). Conversely, the increase in the intensity of the optical band around 1.8 eV observed in the $c$-axis difference spectra upon cooling below $T_C$ has no obvious counterpart in the corresponding $b$-axis spectra. Finally, the optical  response modification at the lowest energies, which is attributable to the shift of the absorption edge below $T_C$, is again most pronounced in $b$-axis polarization. This shift is much less pronounced than the high-temperature changes shown in the left panel of Fig. 12, but its direction is opposite.

Figure 14 shows the temperature dependence of the
$\epsilon_2$ amplitudes at the main peak positions,
1.8 and 2.85 eV, in $b$-axis polarization. In Fig. 15, $\epsilon_1$
and $\epsilon_2$ of the 1.8 eV peak, measured
in $c$-axis polarization slightly off-resonance, are plotted as a function of temperature. Two temperature regimes can be distinguished. The temperature evolution at high temperatures is presumably mainly the result of lattice anharmonicity, which leads to a broadening and shift of spectral features with increasing temperature. Remarkably, a crossover then occurs at $\sim 100$ K, far above $T_C$. Below the crossover point, the temperature dependence of the SW at both peak positions changes sign. (Note that in
$b$-axis polarization, the $\epsilon_2$ amplitude of the 1.8 peak
exhibits nonmonotonic and more complicated behavior, mainly due to
the shift of the absorption edge to lower energies.) Within our experimental
accuracy we are not able to resolve any anomaly at $T_C$. It is noteworthy that the crossover temperature in the optical spectra roughly coincides (within the experimental error) with the onset of deviation of the inverse susceptibility
from the Curie-Weiss behavior discussed in Section II.C. 

Figure 16 shows    the associated spectral weight changes,
$\Delta N_{eff}$(15 K, 55 K), in  $b$- and $c$-axis polarizations, where we also plot  $b$-axis  $N_{eff}(\nu)$ spectrum at 15 K, where 
$N_{eff}(\nu) = \frac{2m}{\pi e^2 N} \int^{\nu}_{\nu_0}\sigma_1(\nu^{'})d\nu^{'}$.
One can notice that $b$-axis spectral weight increases
below $T_C$: (i) at low energies below 1.5 eV, due to the shift of the absorption edge,   and (ii) in the spectral range from 2.2 to 3.5 eV, where it    can be associated with a total spectral weight gain of the optical band at $\sim$ 2.85 eV. In both cases (i) and (ii), one can estimate the SW gain due to  FM ordering, $\Delta N_{eff}$(15 K, 55 K), by approximately the same amount of $\sim$ 0.001. At higher energies, we relate the spectral weight decrease in the spectral range from 3.5 to 4.4 eV to the  optical band around 3.95 eV, however, it is only  weakly displayed. $C$-axis spectral weight changes primarily in the spectral range from 1.5 to 2.3 eV, where it can be associated with a total spectral weight gain of the optical band around $\sim$ 1.8 eV. One can notice that, in the overall, the spectral weight changes in $b$-axis are much more pronounced than those in $c$-axis.  

Based on our observations, we now discuss the origin of the
optical transitions in YTiO$_3$. As discussed in Section III.D.2
above, the temperature dependence of the SW allows us to single
out the contribution from intersite $d_i^1d_j^1$ $\longrightarrow$
$d_i^2d_j^0$ charge excitations to the optical response. Such
transitions lead to four different excited states: a high-spin
(HS) $^3T_1$ state at energy $U^*-3J_H$, two degenerate low-spin
(LS) states $^1T_2$ and $^1E$ at energy $U^*-J_H$, and a LS state
$^1A_1$ at energy $U^*+2J_H$ \cite{Khaliullin,Olesh}. Here $U^*$
is the effective Coulomb repulsion of the two electrons with
opposite spins on the same $t_{2g}$ orbital, and $J_H$ is the Hund
interaction. An ellipsometry study of the LaTiO$_3$ single crystal
(characterized in  \cite{Fritsch}) has revealed that all optical bands are shifted to lower energies by $\sim$ 0.5 eV compared to YTiO$_3$, presumably due to the screening effects. Based on this observation and
the estimated value of the  effective on-site electron correlation
energy $U_{dd}$ $\sim$ 4 eV in LaTiO$_3$ \cite{Higuchi}, we
estimate $U^*$ $\sim$ 4.5 eV in YTiO$_3$. The Hund's coupling
constant is expected to be in the range 0.5--1 eV. (Note that the
free-ion value is 0.59 eV \cite{Gri61}). The optical transition
around 2.85 eV ($\pm$ \ 0.1 \ eV), which exhibits a SW increase in $ab$ polarization
at low temperatures, is therefore naturally assigned to a HS ($^3T_1$) transition, where the valence electron is transferred to an
unoccupied $t_{2g}$ orbital on the neighboring Ti site with a
parallel spin. As the data are noisier at higher energies, the
assignment of the band at 3.95 eV ($\pm$ \ 0.1 \ eV) is more difficult. With this caveat, it can be reasonably attributed to the lowest-lying LS
transition, based on the observation of a SW decrease upon FM spin
ordering. With the resulting tentative values of $U^*$ = 4.5 $\pm$ \ 0.2 \ eV and $J_H$ = 0.55 eV $\pm$ \ 0.1 eV, the next LS transition is expected at $\sim$ 5.6 eV, at the high-energy limit of our optical measurements. In our earlier optical study of LaMnO$_3$ we identified a HS state of $^6A_1$ symmetry at the energy $U^*-3J_H+\Delta_{JT} \sim$ 2.0 eV, where $\Delta_{JT}$ is the Jahn-Teller splitting energy of the $e_g$ levels \cite{Kovaleva}.   The estimated parameters $U^* \sim$ 2.8 eV, $J_H \sim$ 0.5 eV, and $\Delta_{JT} \sim$ 0.7 eV yielded a good description of the observed spectra. This suggests a lower value of energy for the HS state multiplet in LaMnO$_3$ than in YTiO$_3$.   As the values of $U$ (and $J_H$) are supposed to follow certain semiempirical rules and are lower for early transition metal ions than for Mn or ions with almost filled $3d$ shells \cite{Sawatzky,Mizokawa1}, there is a controversy regarding the estimated $U$ values in YTiO$_3$ and LaMnO$_3$. However, there is a good agreement between the $U$ values estimated for the Ti oxides from photoemission spectra \cite{Bocquet}, $t_{2g}$ resonant soft x-ray emission spectroscopy (SXES) \cite{Higuchi}, $ab-initio$ predictions \cite{Aryasetiawan}, and from the present optical study. From this follows that environmental factors (covalency, polarisation, etc.),  which are important in many oxide
systems  \cite{Stoneham1}, are essential to the understanding the
electronic correlations in LaMnO$_3$, in particular, optical transitions and electronic photoinization spectra \cite{Kovaleva1}.

This assignment suggests that the optical band at energy  2
eV (Figs. 9-12) is dominated by transitions of different origins. 
One can see a noticeable shift of the absorption edge to higher energies with decreasing temperature in the left panel of Fig. 12. Theories of
band-tail absorption ascribe this behavior to electrons
interacting with a spatially random potential due to thermal and
structural disorder \cite{Aljishi}. Interestingly, the absorption
edge shifts in the opposite direction below about 100 K, which
results in a SW gain below 1.5 eV (see Fig. 13 (a) and Fig. 16). It
is possible that this effect could be associated with
spin-polaronic (or exciton-polaronic) states in the band-tail density of states or with contributions from indirect $d-d$ optical transitions. In addition, as we have discussed above, weakly dipole-allowed $p-d$ transitions \cite{Zenkov} and intra-site (CF) $d-d$ transitions can contribute here. Further experimental and theoretical  studies are required to estimate different
contributions to the  optical band at energy 2 eV.    

We now focus on the absolute SW gain of the (HS) $^3T_1$-state $d-d$ transition
centered around 2.85 eV due to  $spin \ ordering$ - between
the high-temperature limit (HT) - disordered spins - and the low-temperature limit (LT) - FM ordered spins,  as well as the SW gain at low
temperatures.  Via the optical sum rule in the tight-binding approximation, one finds $N_{eff}=(ma_0^2/\hbar^2)K$, where $K$ is the kinetic energy
associated with virtual charge excitations. The contribution of
the (HS)$^3T_1$ excitation to the kinetic energy $K$ can be
estimated from a related term in the superexchange energy, $K =
-2\langle H_{SE}(^3T_1)\rangle$. For the bond along the
$\delta(=a,b,c)$ direction, $H^{(\delta)}_1 =
\frac{2t^2}{U-3J_H}(\bar S_i\cdot \bar
S_j+3/4)(A^{(\delta)}_{ij}-\frac{1}{2}{n_{ij}^{(\delta)}})$, where
the orbital operators ${A^{(\delta)}_{ij}}, n^{(\delta)}_{ij}$
depend on the direction of the $\langle ij\rangle$ bond and $t$ is
the electron transfer amplitude \cite{Olesh,Khaliullin}. Provided
the optical excitation energy of the HS-state at $U-3J_H$ is
specified, the optical spectral weight depends on the transfer
amplitude $t$ and the underlying orbital state.  We estimate the
SW in the framework of the isotropic $F$-type classical orbital
ordering, considered by Ole{\'s} {\it et al.} \cite{Olesh}. The
kinetic energy of the HS-state at $T$ = 0 K is then expressed as
$K^{(\delta)}_1$ = $\frac{2}{3}\frac{4t^2}{U-3J_H}$. At high
temperatures, spin disorder reduces this value by a factor of 1/4.
In the FM state the isotropic exchange constants are determined by
$J^{(\delta)}_{FM}$ = $-\frac{2}{3}\frac{4t^2
J_H}{(U-3J_H)(U-J_H)}$. Using the experimental value of the
exchange constant $J^{(\delta)}$ = $-$3 meV \cite{Ulrich}, and
conservatively considering a wide range of Hund coupling constants
0.5 eV $\leqslant J_H\leqslant$ 1 eV, the transfer amplitude is
estimated in the range 120 meV $\lesssim t\lesssim$ 160 meV, in
reasonable agreement with {\it ab-initio} calculations
\cite{Pavarini}. The resulting estimate of the SW in the
paramagnetic state, 0.022 $\lesssim N^{(\delta)}_1 (T\gg
T_C)\lesssim$ 0.034, is a factor of 2-3 lower than the
$b$-axis experimental value listed in Table I. This estimate gives credence
to our contention that inter-site $d-d$ transitions contribute
substantially to the optical SW in this energy range.
The same analysis yields a theoretical value of 0.008 $\lesssim
\Delta N_1^{(\delta)}$ (LT, HT) $\lesssim$ 0.012 for the SW gain
of the (HS)$^3T_1$ transition due to $spin \ ordering$. The
experimental value in $b$-polarization is $\Delta
N^{ab}_{HS}$(15K, 55K) $\simeq$ 0.002, due to a total contribution from the
direct $d-d$ transitions at $\sim$ 2.85 eV and the indirect $d-d$ transitions
(or spin-polaronic states) at the absorption edge (see Fig. 16). 
As the onset of the SW transfer is about 100 K, and the transfer is only about half complete at 55K, the difference between experimental and
theoretical values is again about a factor of 2-3.

The absence of any noticeable temperature-dependent SW change of the
band at 2.85 eV along the $c$-axis (Fig. 13b) disagrees with the
simple analysis presented above.  One might be tempted to attribute
this discrepancy to a strong anisotropy of the superexchange parameters
predicted by various theoretical calculations \cite{Pavarini,Mochizuki}. However, according to our optical data, the spectral weight of the low-energy
optical bands along the $c$-axis is  only a factor of  1.5-2
smaller than in the $ab$-plane (Table I). Moreover, a direct
determination of the exchange parameters by inelastic neutron
scattering is inconsistent with this prediction \cite{Ulrich}. 
  
To summarize the discussion, we note that the overall magnitude of
the SW of the optical band at 2.85 eV and its temperature
dependence in $ab$-polarization are in reasonable agreement with
the predictions of a simple superexchange model, in analogy to
LaMnO$_3$. However, the model overestimates the SW gain of the (HS)$^3T_1$ transition due to $spin \ ordering$ by a factor of 2-3.
In addition, the anisotropic temperature dependence of
the optical SW  indicates that an understanding of the
optical conductivity of YTiO$_3$ requires mechanisms that go
qualitatively beyond this simple model. These observations can be naturally understood as a consequence of a possible Ti $t_{2g}$
orbital rearrangement between 55 K and 15 K. Then, the lowering of the optical spectral weight and the anisotropy come through selection rules of the modified ground state wavefunction. 

This may imply  an important $qualitative$ difference between LaMnO$_3$ and
YTiO$_3$.  Almost undistorted TiO$_6$ octahedra in  YTiO$_3$ indicate a small Jahn-Teller instability, resulting from the weak  coupling between the two nearly-degenerate electronic ground states. The Jahn-Teller instability may disappear if there is sufficiently large splitting of the nearly-degenerate levels by some perturbation, in particular by the spin-orbit interactions \cite{Opik,Stoneham}. The orbital degrees of freedom may be essentially quenched in the magnetically ordered state, when the ground multiplet experiences splitting into two doublets, with concomitantly, a partial uncoupling of the spin and orbital moments. Then the temperature behavior  in such a system is largely determined by the interplay between the  spin-orbit coupling and the Jahn-Teller effect. In addition, the joint spin-and-orbital fluctuations, which develop at finite
temperatures, can contribute significantly at  $T > T_C$ \cite{Khaliullin2}.
In such a case, the establishment of the spin-spin correlations in approaching the magnetic transition will give rise to complex $spin-orbit-lattice$ effects, peculiar to the systems with orbital degrees of freedom.   By contrast, the orbital degrees of freedom are quenched in LaMnO$_3$ by a large static Jahn-Teller distortion, and with the use of the same approach we have shown that the redistribution of the optical spectral weight is in a good agreement with a superexchange model that attributes these shifts to the temperature dependent correlations between Mn spins \cite{Kovaleva}. 

Finally, in Fig. 17 we show temperature dependence of the $b$-axis
optical conductivity investigated  on the aged surface of our
YTiO$_3$ single crystal. In contrast to the temperature
dependencies measured on freshly polished surfaces (see Fig. 13),
a sizable temperature dependence is apparent in these spectra.
With decreasing temperature, the optical SW is transferred from
higher energies to the low-energy band around 1.9 eV. This
temperature effect, measured on the aged surface of our YTiO$_3$
single crystal, is very similar to the observations recently
reported by G\"{o}ssling {\it et al.} \cite{Gruninger}. However,
one can hardly ascribe this temperature effect to spin-spin
correlations, because the temperature changes are observable over
a wide temperature range from 300 K to 15 K, and the
high-temperature changes are much more pronounced than those that
might be associated with the onset of correlated spin fluctuations
at low temperatures. In addition,  Fig. 7 shows that the optical
conductivity spectra that exhibit a noticeable high-temperature
effect, differ from those of measured on freshly polished
surfaces. As discussed in Section III.C, we associate this
phenomenon with the presence of defects of the titanium oxides, with the
valency of Ti increased above 3$^+$, and nearby extra oxygen (interstitials)
in the near-surface layer. The observed redistribution of the SW may then be associated with localized states at the edge of the $p$ and $d$ electronic bands.

\section*{IV. CONCLUSION}

The detailed ellipsometric study presented here shows that oxygen defects can strongly modify the optical response in the near-surface region probed by the light beam. Extreme care must therefore be taken to maintain oxygen stoichiometry at the surface under investigation. The intrinsic optical response of YTiO$_3$ is broad, weakly anisotropic, and weakly temperature dependent. Nonetheless, using a careful study of the temperature dependence in combination with a dispersion analysis, we were able to identify an optical band at 2.9 eV whose energy, absolute spectral weight, and temperature dependence is consistent with the intersite $d_i^1d_j^1$ $\longrightarrow$ $d_i^2d_j^0$ HS-state ($^3T_1$) optical transition. This indicates that $d-d$ transitions constitute a significant part of the optical response below the onset of dipole-allowed transitions between oxygen $p$- and titanium $d$-states at about 5 eV. However, this analysis accounts for only part of the optical response 
 of YTiO$_3$. In particular, the origin of an intense optical band around 2 eV remains unclear. This indicates that other mechanisms, such as weakly dipole-allowed $p-d$ transitions or spin-polaron (or exciton-polaron) excitations, can contribute here significantly to the optical spectra. A more elaborate approach is required to obtain a quantitative description and verify the assignment of this optical band. 
   
Our experiments also uncovered an important {\it qualitative} difference between LaMnO$_3$ and YTiO$_3$. In LaMnO$_3$, the spectral weight of inter-site $d-d$ transitions exhibits a strong anomaly upon heating above the N\`eel temperature, followed by a more gradual temperature dependence due to short-range spin correlations at higher temperature. In YTiO$_3$, the optical spectral weight evolves smoothly through the onset of magnetic long-range order, with no discernible anomaly at the Curie temperature $T_C = 30$ K. However, a distinct anomaly is observed at about 100 K, far above $T_C$ but coincident with a deviation of the uniform magnetization from mean-field Curie-Weiss behavior. This observation appears inconsistent with models in which the orbital degeneracy is completely quenched by lattice distortions, as it is in LaMnO$_3$, and gives credence to models that allow for a temperature-dependent orbital rearrangement. Our data therefore underscore the role of YTiO$_3$ as an interesting model system for the complex interplay between spin, orbital, and lattice degrees of freedom.

\section*{ACKNOWLEDGMENTS}

We thank O. K. Andersen, M. Mochizuki  for fruitful discussions, G. Khaliullin for motivation in this work and  stimulating discussions, Y.-L. Mattis for the support at the IR beam line of ANKA synchrotron
at Forschungszentrum Karlsruhe, and Mr. Wendel for the sample alignment and
preparation.

.

\newpage

\begin{figure}[tbp]
\includegraphics*[width=180mm]{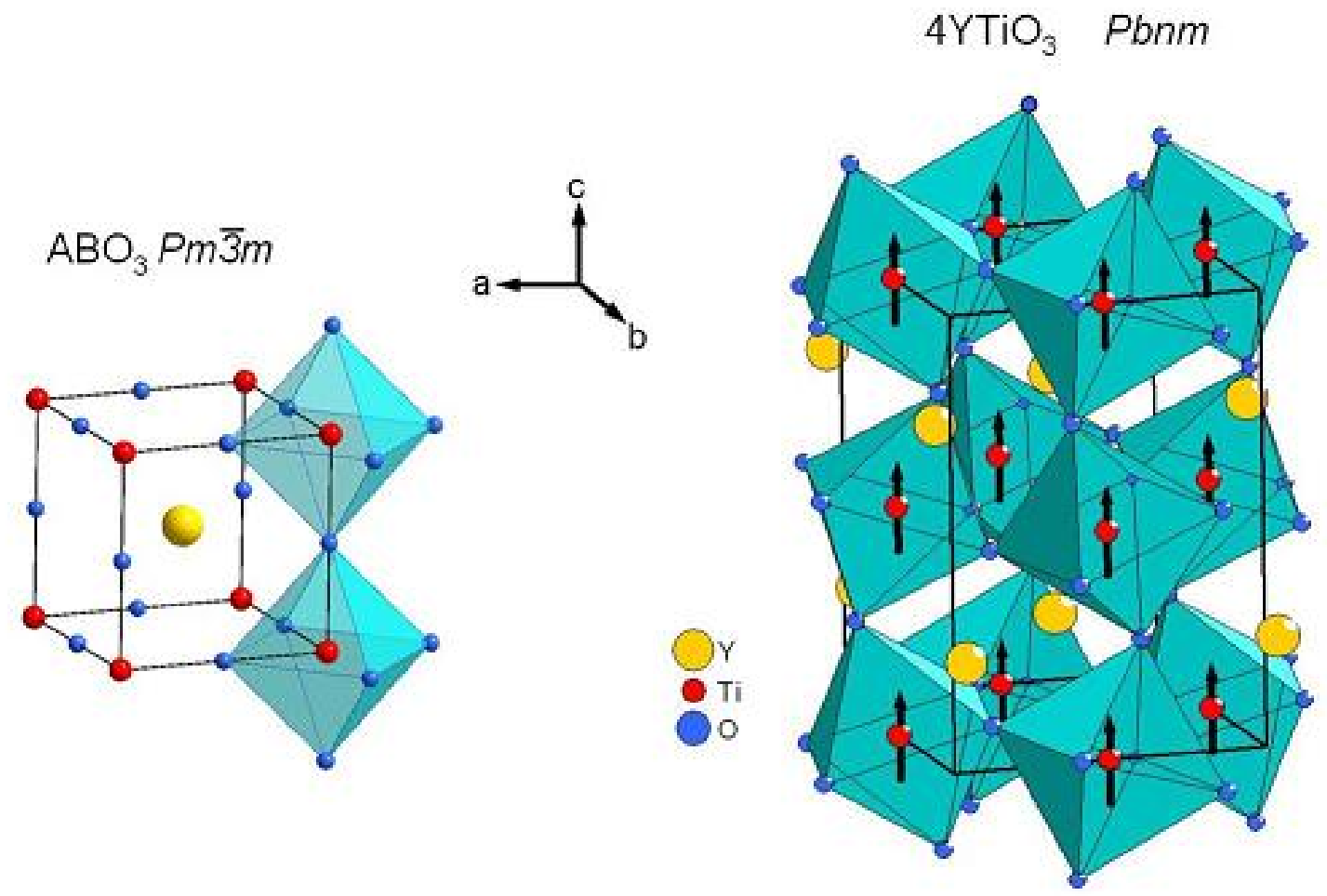}
\caption{(Color) Crystal structures of cubic perovskite oxides of the formula ABO$_3$ (space group $Pm\bar3m$), where A is a rare-earth atom (yellow sphere) and B is a transition-metal atom (red spheres) (left), and related orthorhombic structure of YTiO$_3$ (space group $Pbnm$) with GdFeO$_3$-type distortions, with ferromagnetic alignment of Ti spins shown schematically (right).} 
\label{Fig1}
\end{figure}

\begin{figure}[tbp]
\includegraphics*[width=180mm]{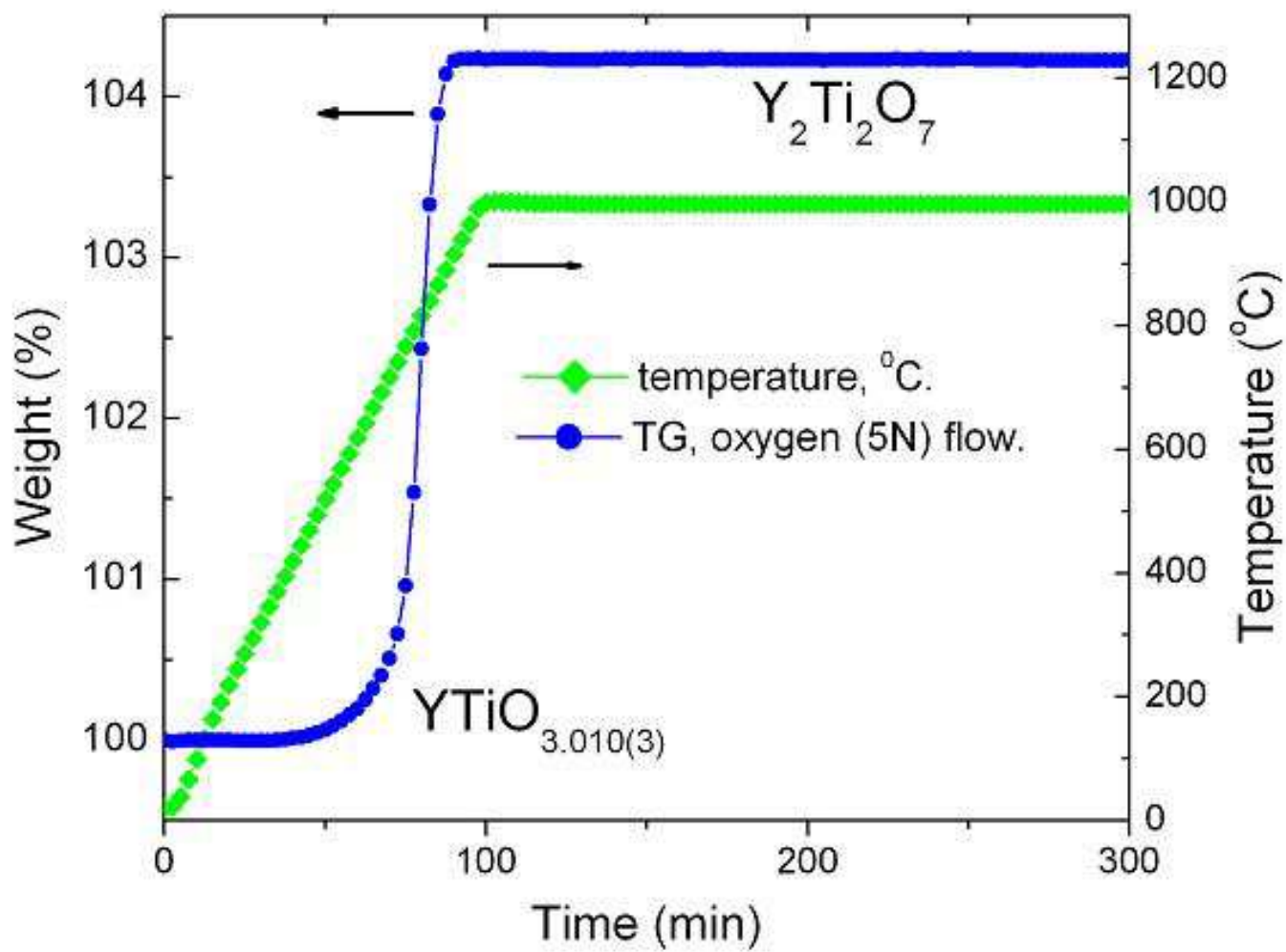}
\caption{(Color online) Weight gain (circles) of YTiO$_3$ sample in oxygen atmosphere versus temperature (squares). As a result of full oxidation of Ti$^{3+}$ to Ti$^{4+}$ Y$_2$Ti$_2$O$_7$ compound is stabilized.}
\label{Fig2}
\end{figure}

\begin{figure}[tbp]
\includegraphics*[width=160mm]{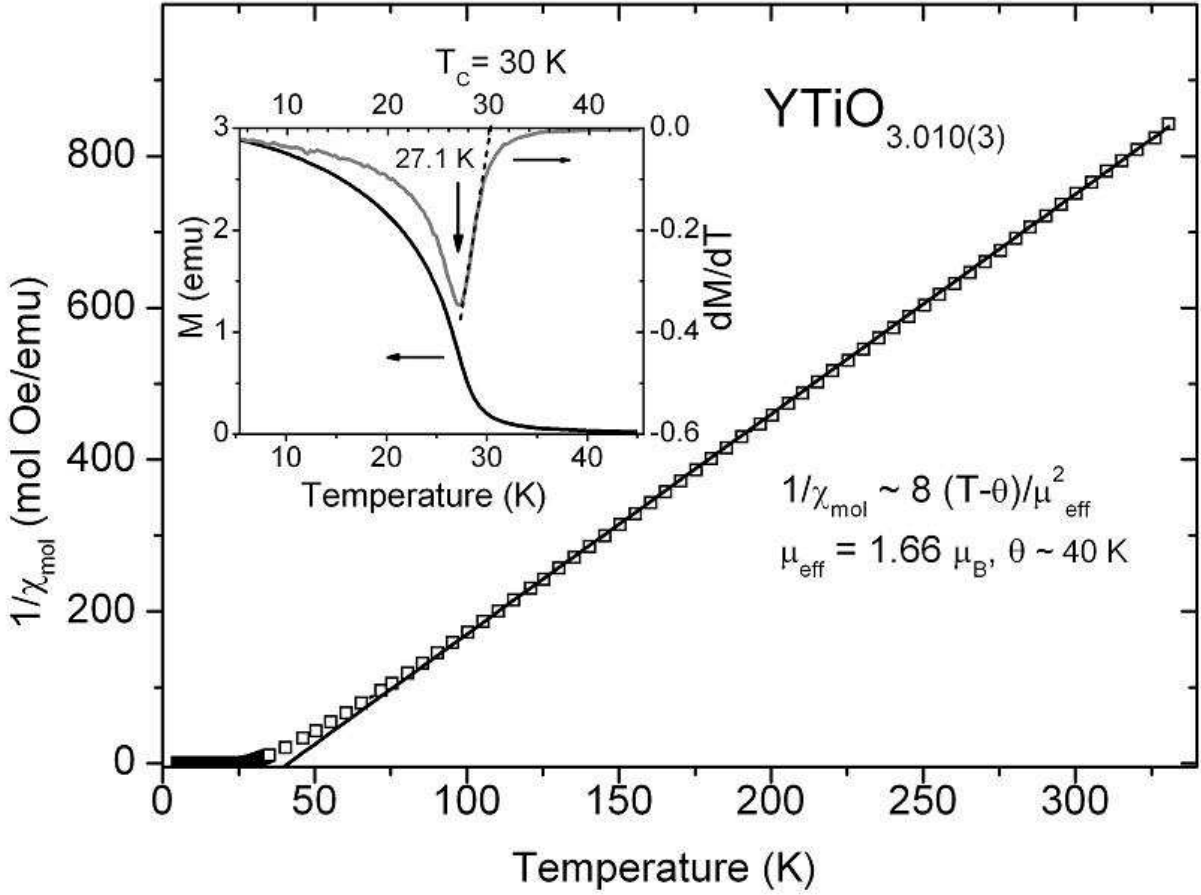}
\caption{Temperature dependence of inverse susceptibility
(squares) of YTiO$_3$ single crystal approximated by Curie-Weiss behavior
at high temperatures (solid line). The inset shows temperature dependence of magnetization in the vicinity of T$_C$ and its temperature derivative
after zero-field cooling in magnetic field of 1000 Oe with {\bf H} $\parallel$ $c$. An inflection point in temperature-dependent
magnetization is observed at 27 K.} 
\label{Fig3}
\end{figure}

\begin{figure}[tbp]
\includegraphics*[width=160mm]{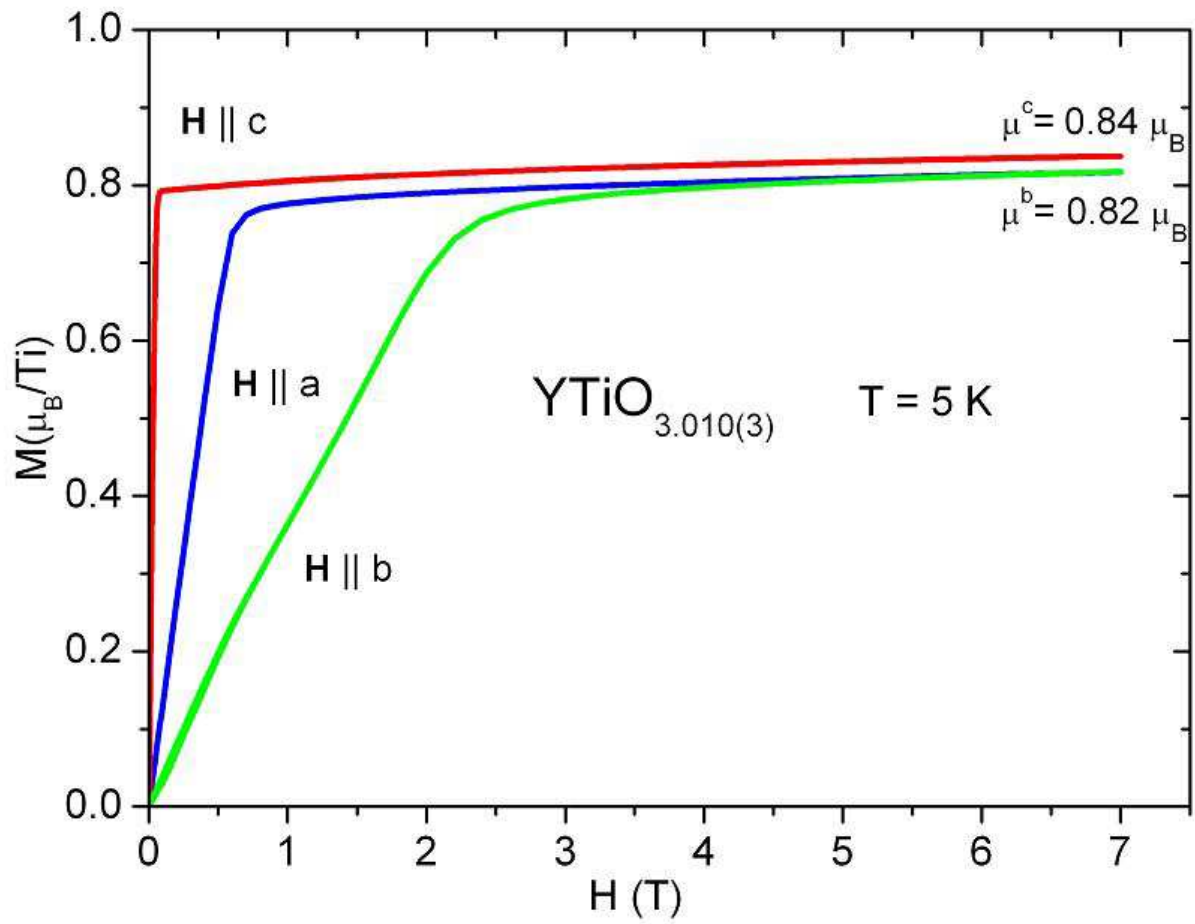}
\caption{(Color online) Magnetization of YTiO$_3$ single crystal measured with magnetic field applied along $a$-, $b$-, and $c$-axes (in $Pbnm$ notation)
at T = 5 K.} 
\label{Fig4}
\end{figure}

\begin{figure}[tbp]
\includegraphics*[width=140mm]{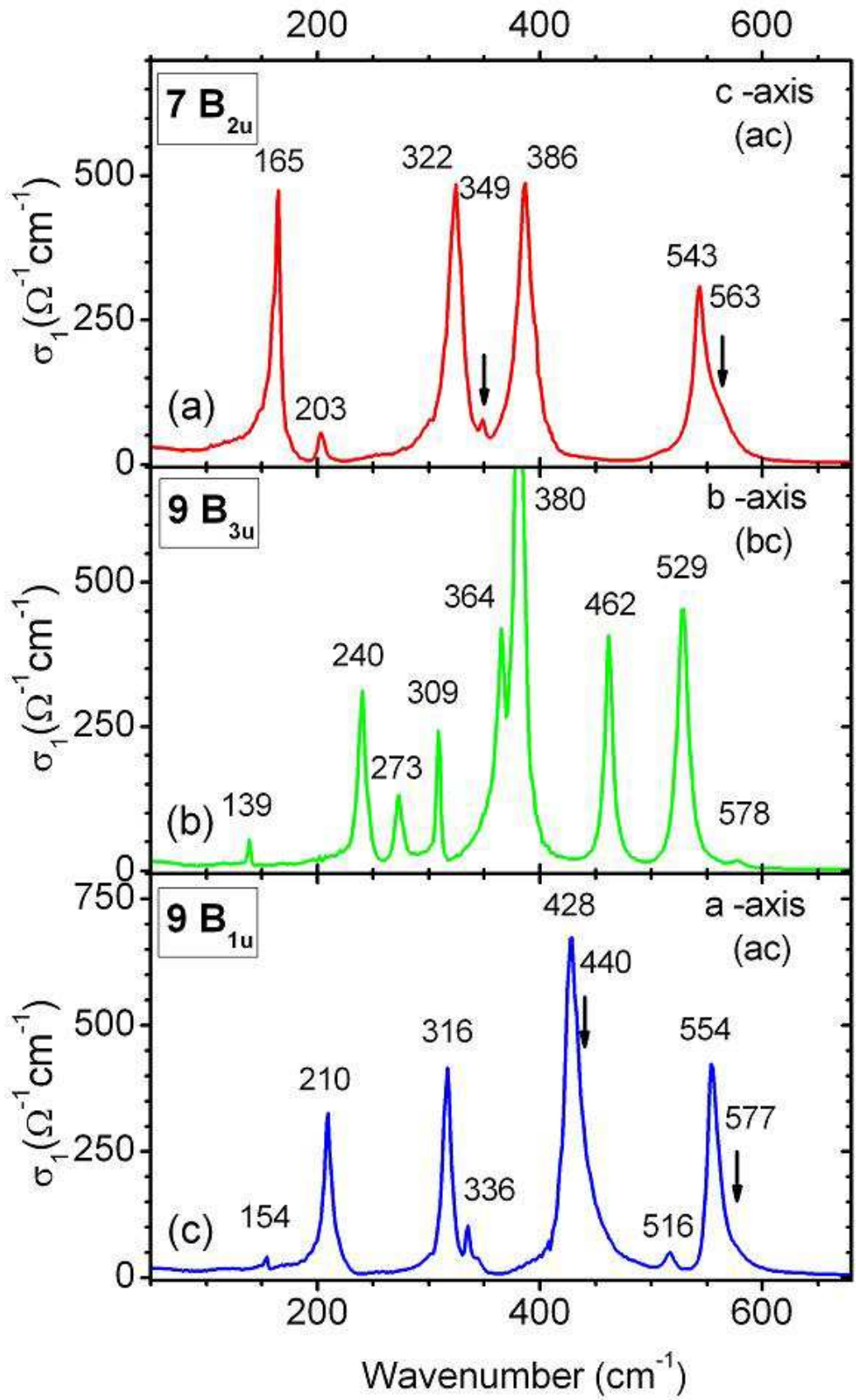}
\caption{(Color online) IR-active optical phonons from polarized reflectivity measurements along (a) $c$-axis, (b) $b$-axis, and (c) $a$-axis at T = 15 K.} 
\label{Fig5}
\end{figure}

\begin{figure}[tbp]
\includegraphics*[width=160mm]{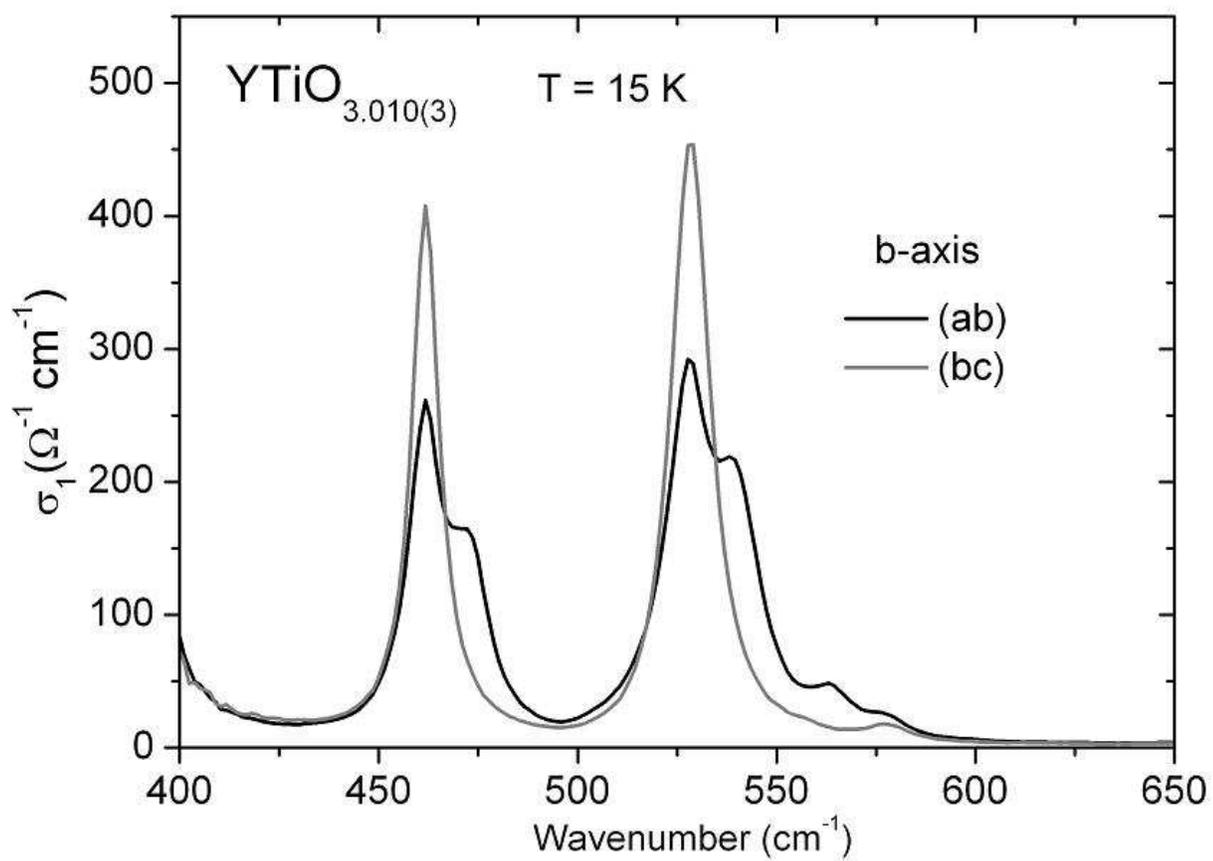}
\caption{Oxygen-dominated $b$-axis optical phonons in YTiO$_3$ single crystal measured on $ab$ and $bc$ surfaces at 15 K.} 
\label{Fig6}
\end{figure}

\begin{figure}[tbp]
\includegraphics*[width=160mm]{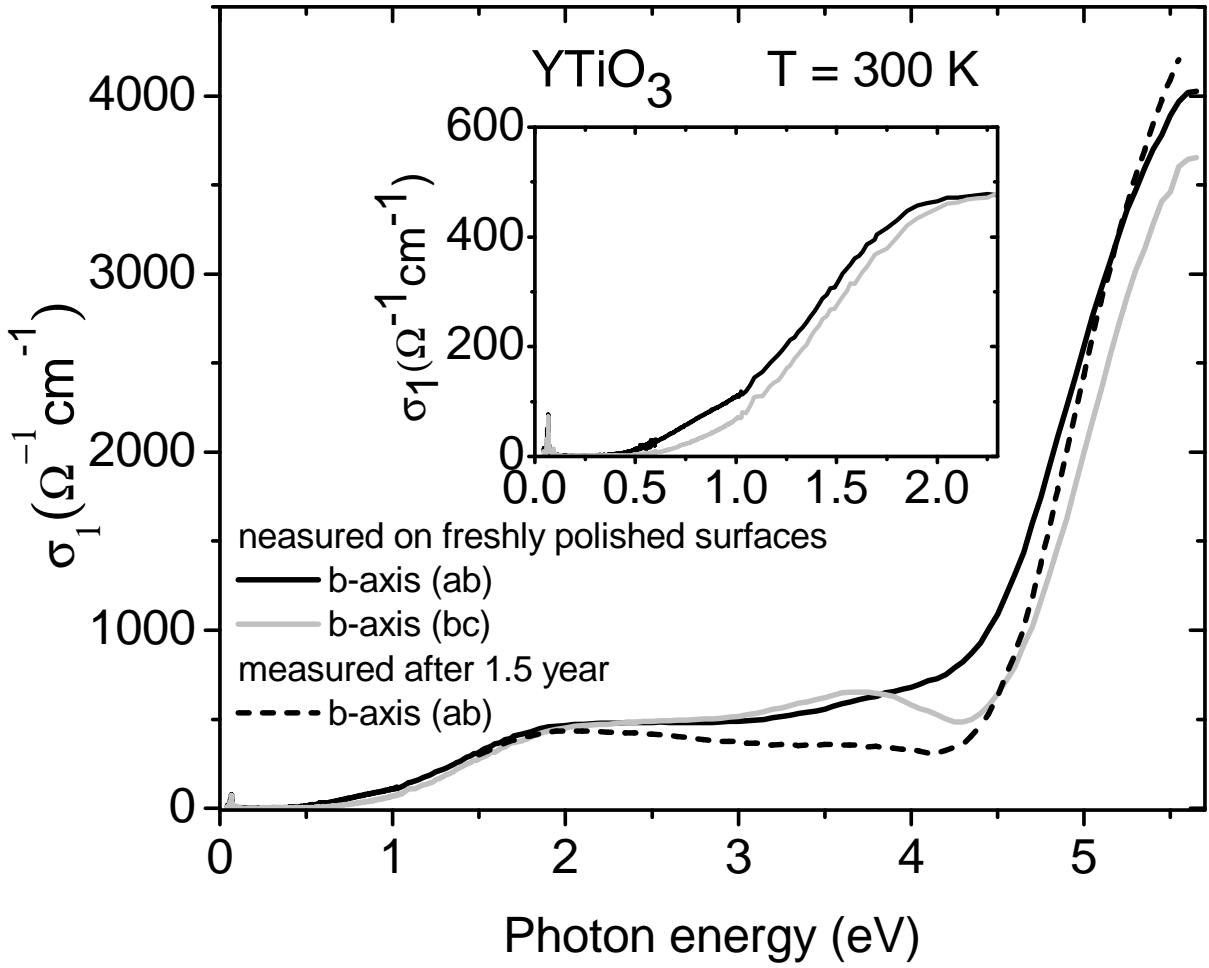}
\caption{$B$-axis optical conductivity spectra measured on freshly polished  $ab$ and $bc$ surfaces and on aged $ab$ surface at T = 300 K. The inset shows details near the absorption edge.} 
\label{Fig7}
\end{figure}

\begin{figure}[tbp]
\includegraphics*[width=170mm]{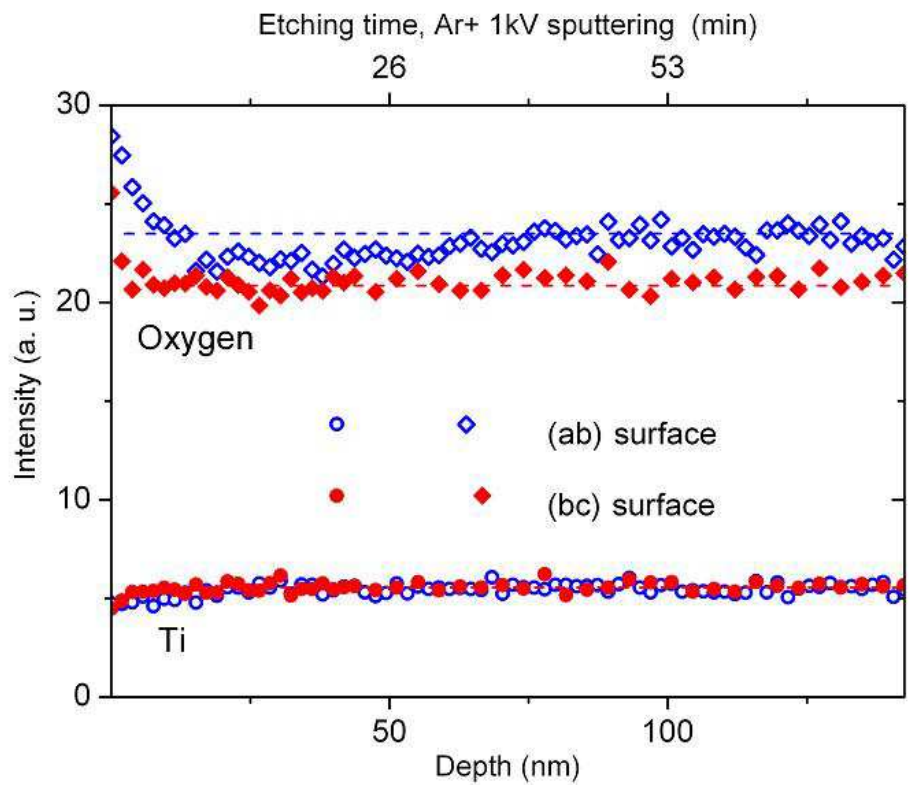}
\caption{(Color online) AES sputter-depth profiles on the $ab$ and $bc$ surfaces in orthorhombic YTiO$_3$ ({\it Pbnm}), associated with (001)- and (110)-perovskite-type surfaces.} 
\label{Fig8}
\end{figure}

\begin{figure}[tbp]
\includegraphics*[width=155mm]{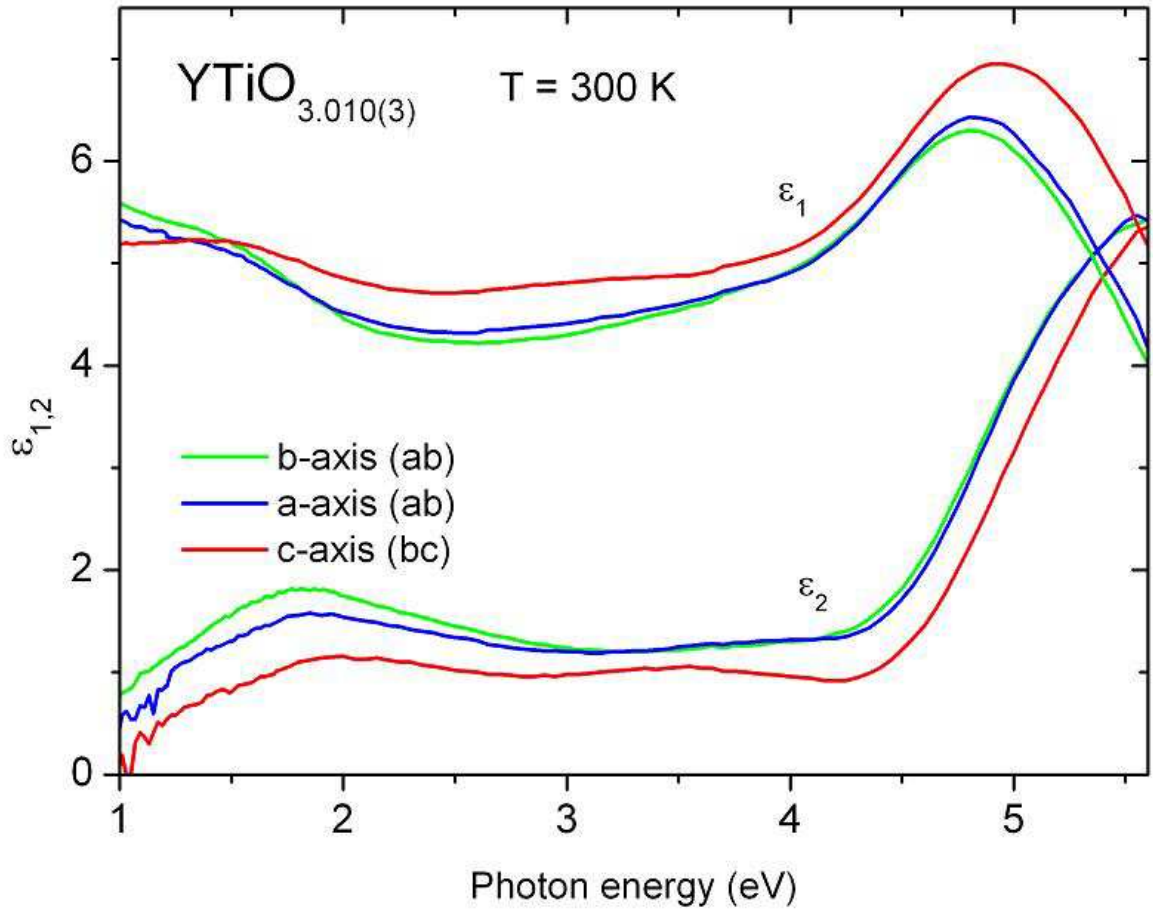}
\caption{(Color) Dielectric function of YTiO$_3$ single crystal in $a$- and $b$-axis polarizations measured on $ab$ plane, and in $c$-axis polarization measured on $bc$ plane at T = 300 K.}
\label{Fig9}
\end{figure}

\begin{figure}[tbp]
\includegraphics*[width=160mm]{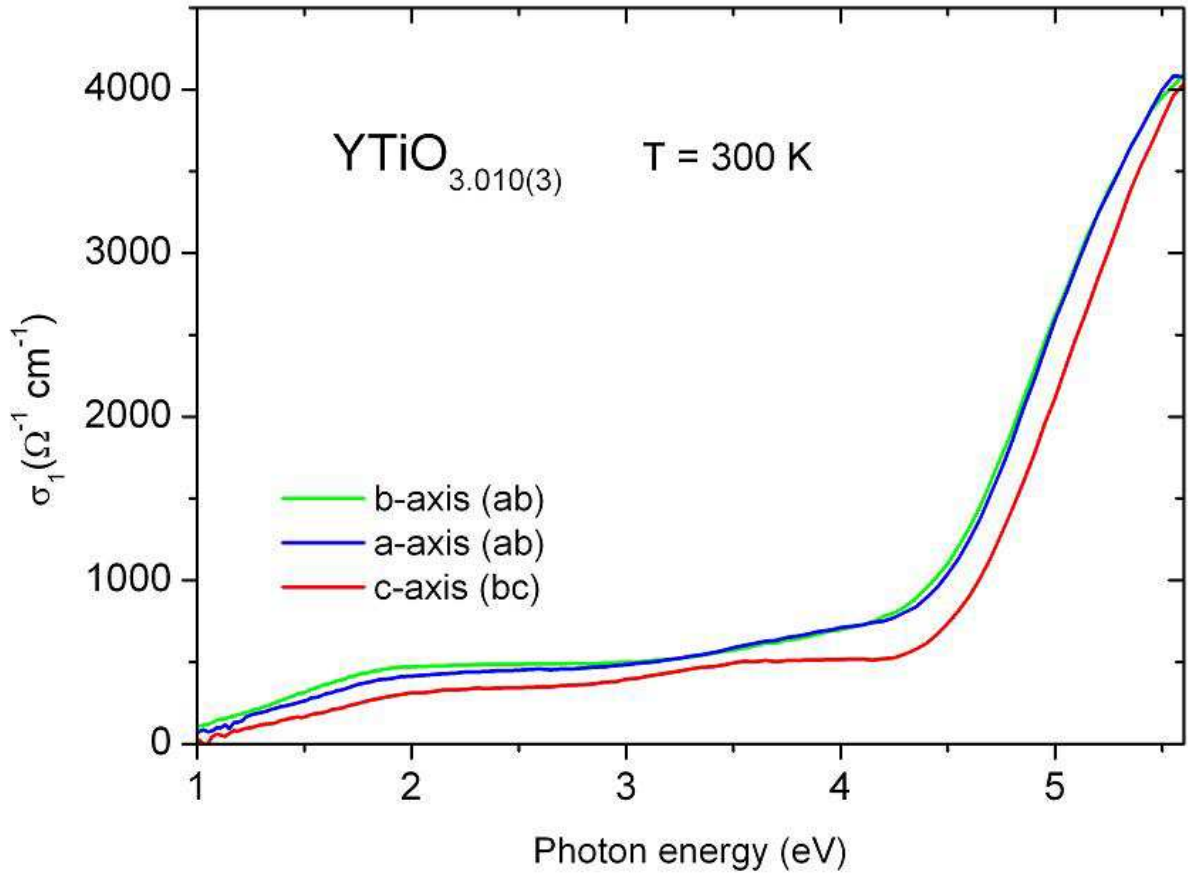}
\caption{(Color) Optical conductivity of YTiO$_3$ single crystal in $a$- and $b$-axis
polarizations measured on $ab$ plane, and in $c$-axis polarization measured on $bc$ plane at T = 300 K.}
\label{Fig10}
\end{figure}

\begin{figure}[tbp]
\includegraphics*[width=155mm]{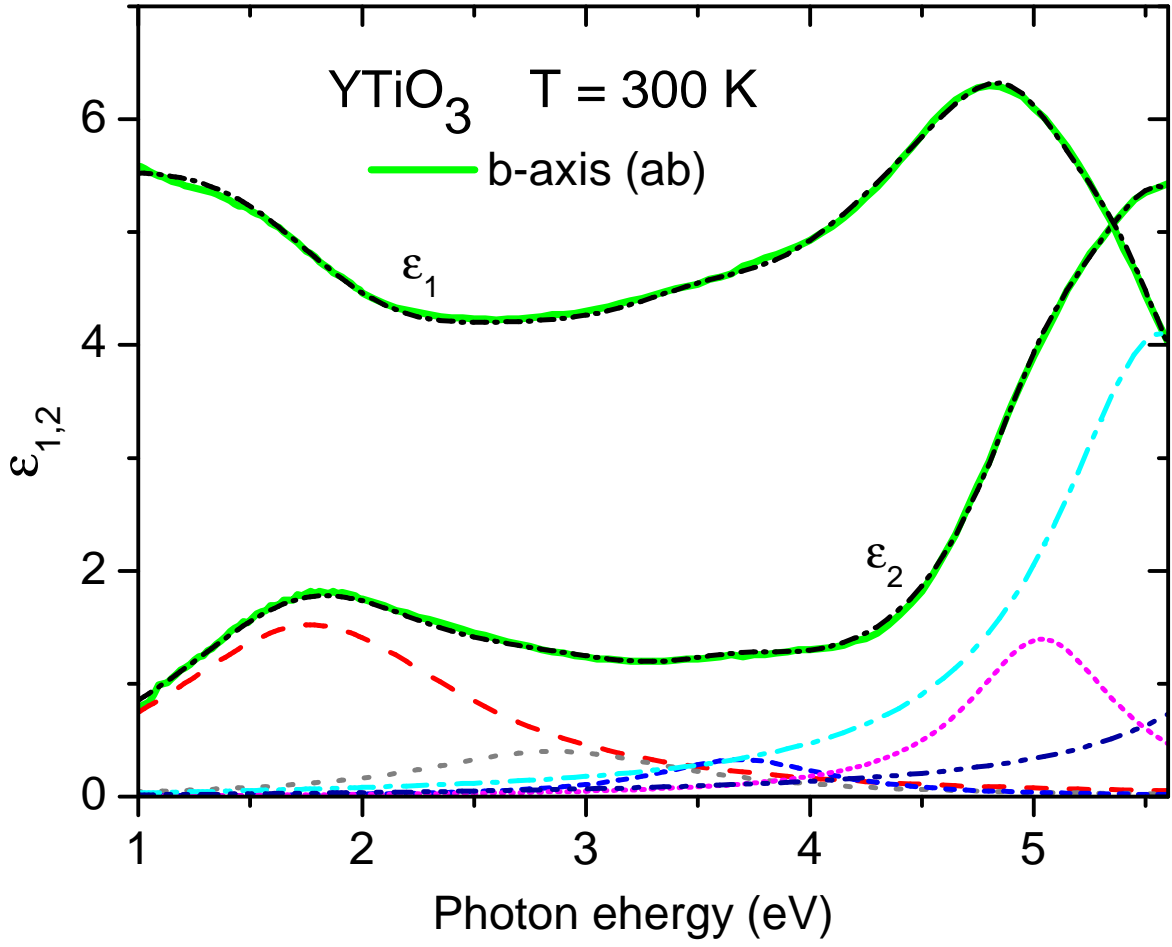}
\caption{(Color online) $B$-axis dielectric response $\tilde\epsilon(\nu)$ at 300 K, represented by total contribution (black dash-dotted curves) of separate Lorentzian bands determined by the dispersion analysis,
as described in the text. Peak energies $\nu_j$, widths $\gamma_j$, and dimensionless oscillator strengths $S_j$, of Lorentzian bands are listed in Table 1.}
\label{Fig11}
\end{figure}

\begin{figure}[tbp]
\includegraphics*[width=180mm]{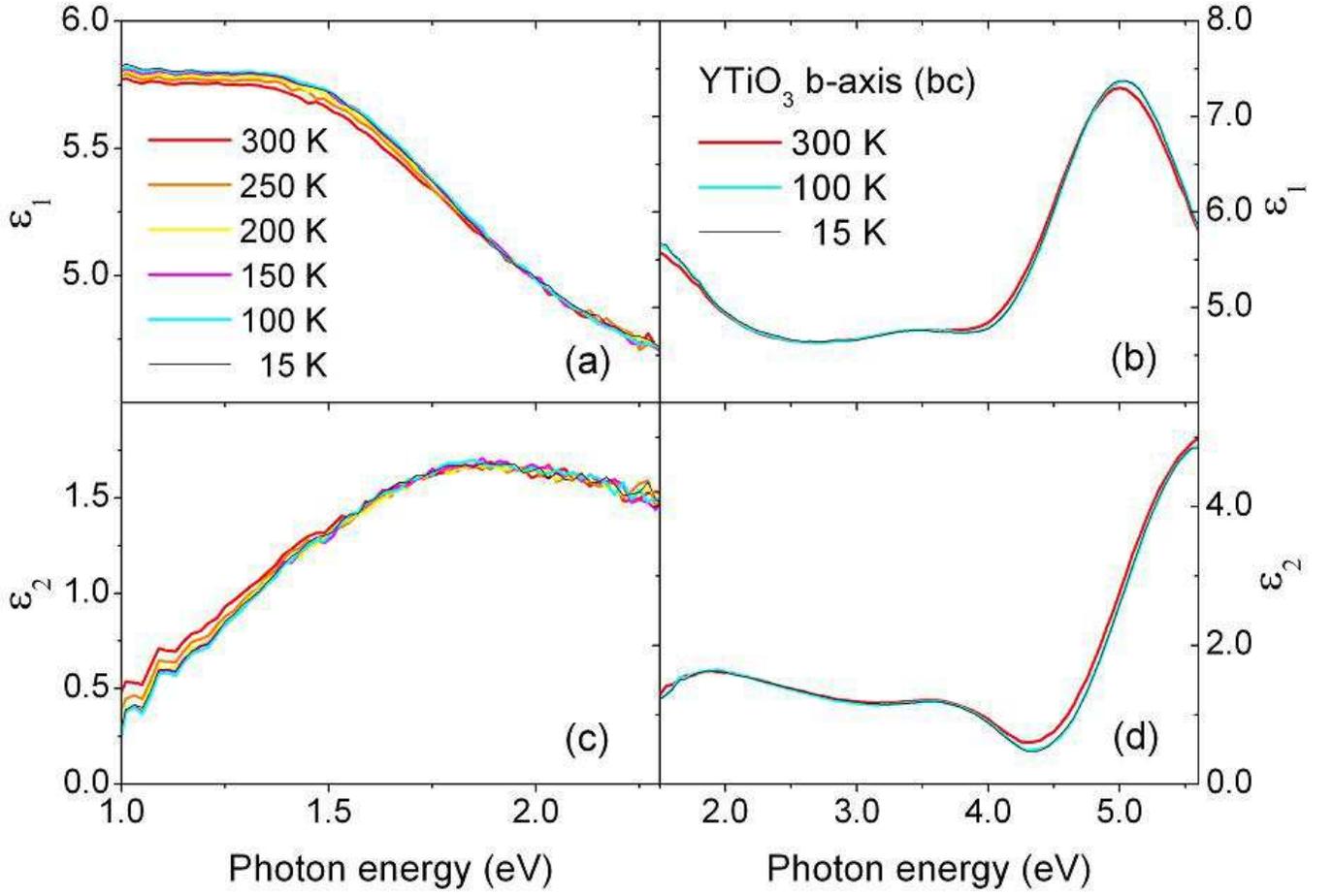}
\caption{(Color) Temperature dependences of: (a) and (b) $\epsilon_1(\nu)$ and (c) and (d) $\epsilon_2(\nu)$ of YTiO$_3$ single crystal measured in $b$-axis polarization on $bc$ plane. Left panels detail changes of complex dielectric response near the absorption edge; right panels show its major changes at higher energies.}
\label{Fig12}
\end{figure}

\begin{figure}[tbp]
\includegraphics*[width=150mm]{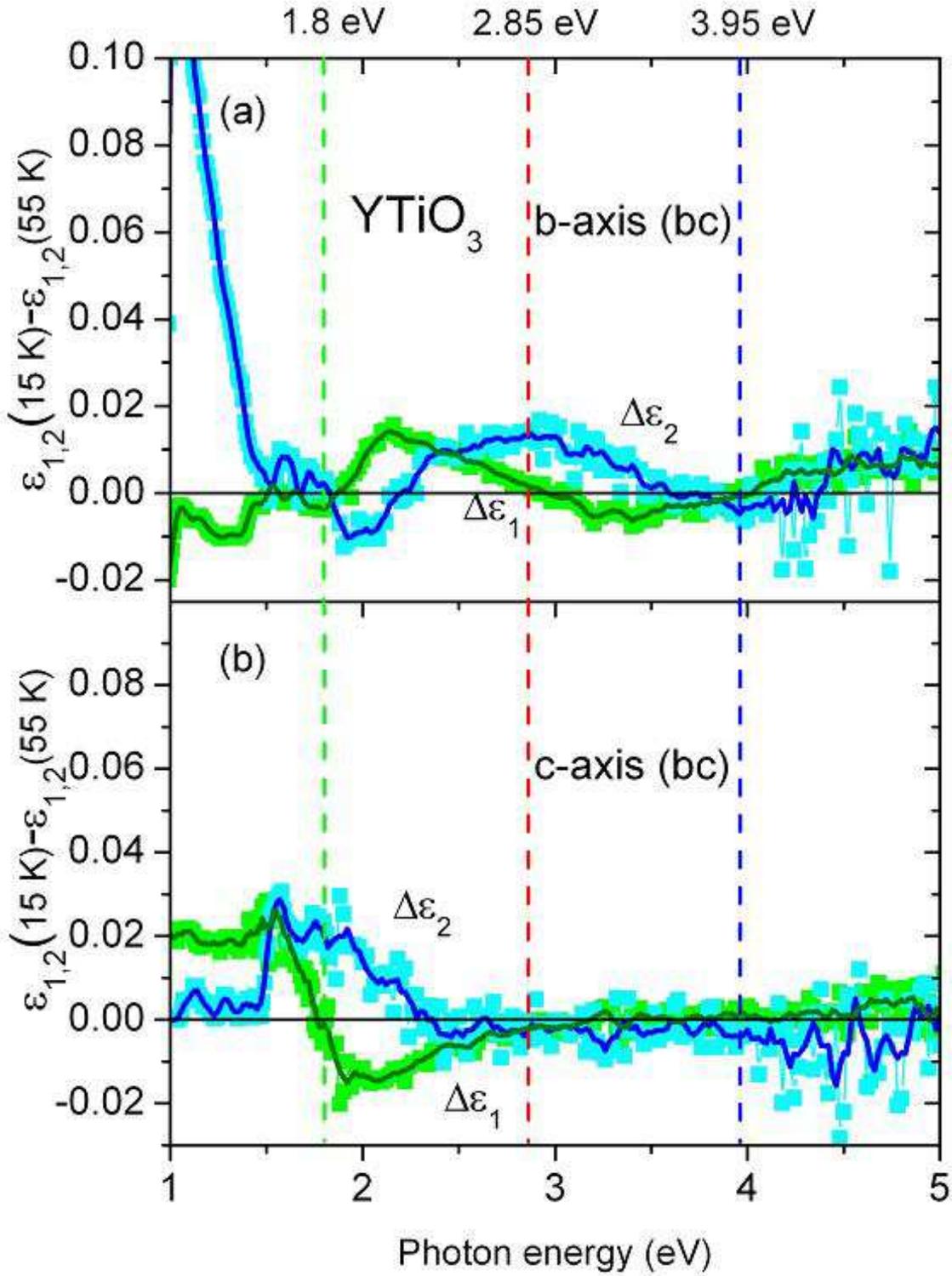}
\caption{(Color) Changes in $b$- and $c$-axis complex dielectric response $\tilde\epsilon(\nu)$ of YTiO$_3$ single crystal between 15 and 55 K (solid curves are result of averaging). Vertical lines mark photon energies at which resonant behavior in $\Delta\epsilon_2(\nu)$ coincides with zero-crossing in $\Delta\epsilon_1(\nu)$ in  $b$- and/or $c$-axis polarization(s).}
\label{Fig13}
\end{figure}

\begin{figure}[tbp]
\includegraphics*[width=130mm]{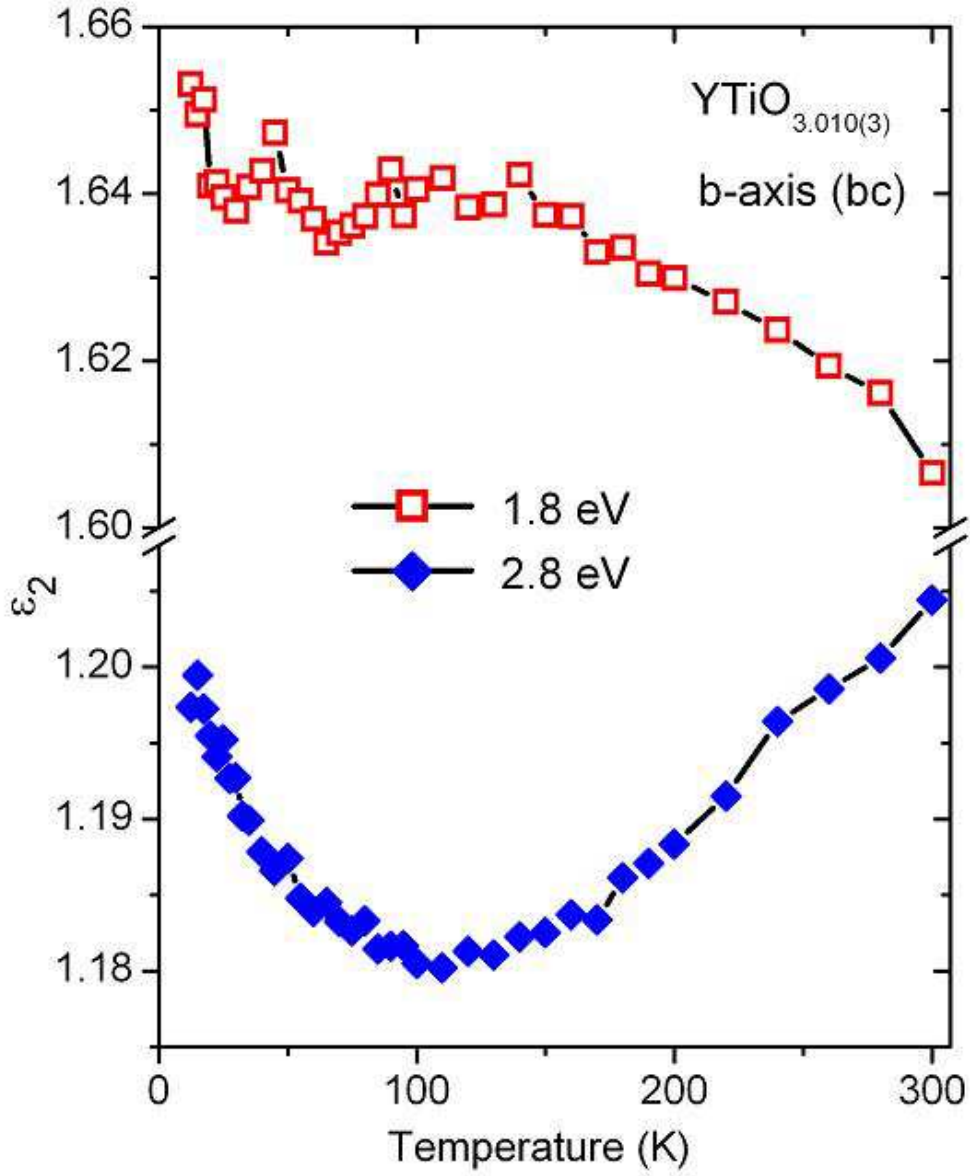}
\caption{(Color online) Temperature dependences of $\epsilon_2$ intensities at the peak positions 1.8 and 2.85 eV in $b$-axis polarization.} 
\label{Fig14}
\end{figure}

\begin{figure}[tbp]
\includegraphics*[width=160mm]{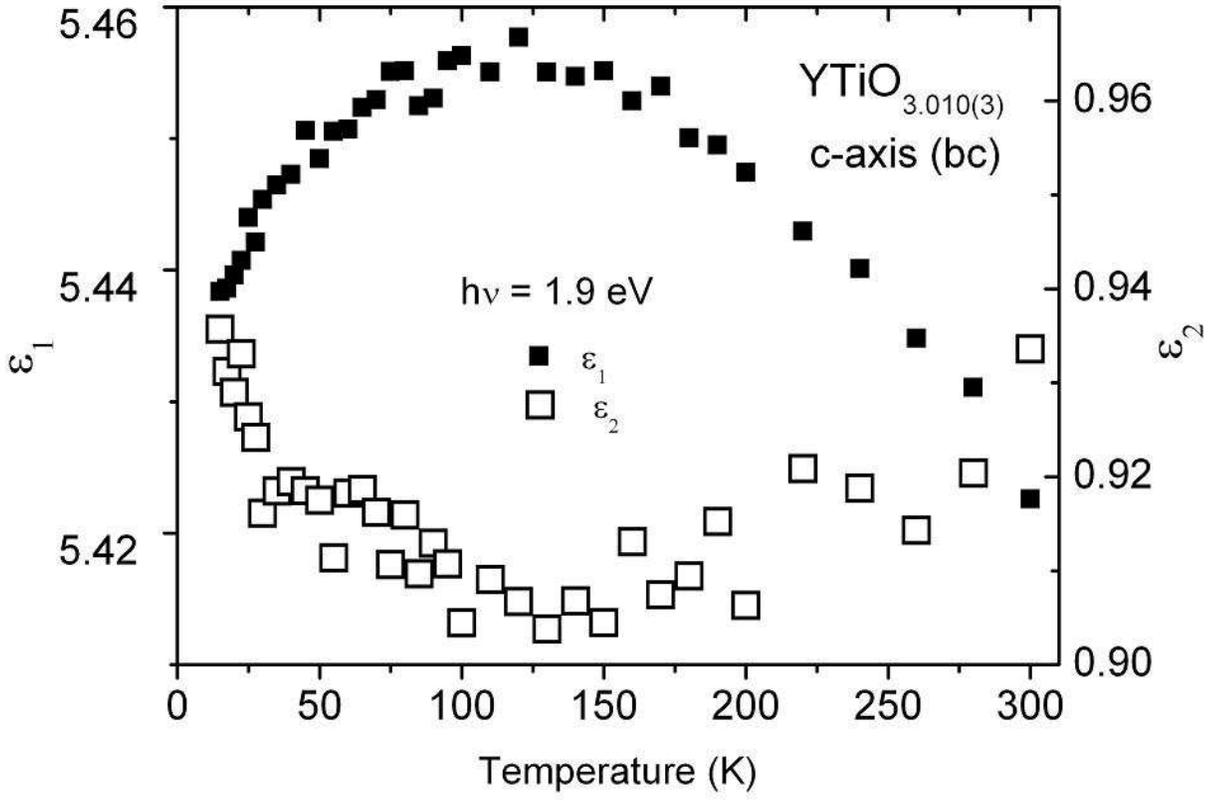}
\caption{Temperature dependences of $\epsilon_1$ and $\epsilon_2$ intensities slightly off-resonance to the peak position at 1.8 eV in $c$-axis polarization.}
\label{Fig15}
\end{figure}

\begin{figure}[tbp]
\includegraphics*[width=160mm]{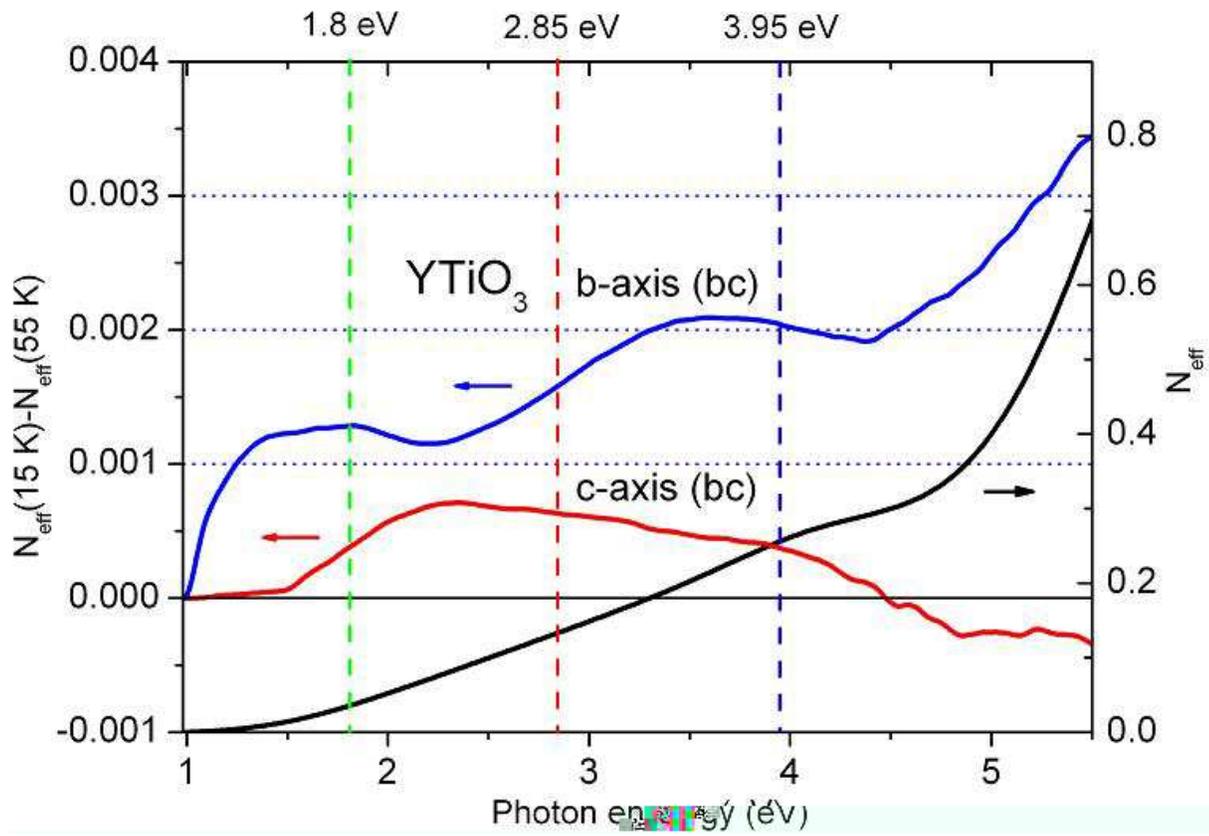}
\caption{(Color online) Spectral weight changes $\Delta N_{eff}(\nu)$ between 15 and 55 K in $b$- and $c$-axis polarizations of YTiO$_3$ single crystal
and $b$-axis $N_{eff}(\nu)$ spectrum at 15 K. Vertical lines
mark peak positions of temperature-dependent optical bands in $b$- and/or $c$-axis polarization(s).}
\label{Fig16}
\end{figure}

\begin{figure}[tbp]
\includegraphics*[width=160mm]{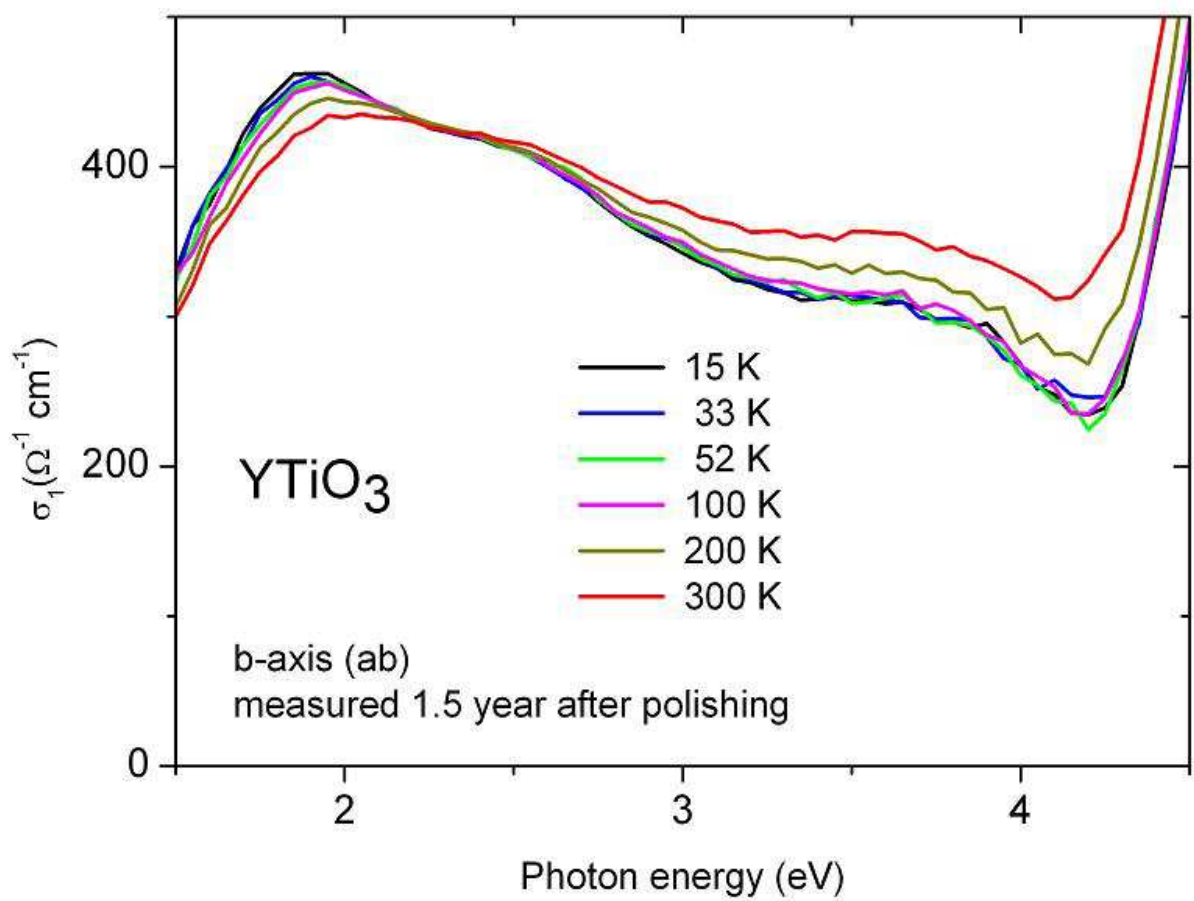}
\caption{(Color) $\sigma_1(T)$ measured on the aged surface of YTiO$_3$ single crystal.}
\label{Fig17}
\end{figure}

\end{document}